\documentclass[aps,prx,twocolumn,showpacs,psfig,superscriptaddress,longbibliography]{revtex4-2}

\usepackage{times}
\usepackage{graphicx}
\usepackage{float}
\usepackage{latexsym,amsmath,amssymb,bm,euscript}
\usepackage{color}
\usepackage{subfigure}
\usepackage{epstopdf}
\usepackage[colorlinks=true,linkcolor=blue,citecolor=blue]{hyperref}
\usepackage{soul}
\usepackage[normalem]{ulem}
\usepackage{mathrsfs}
\usepackage{amsmath}
\usepackage{lettrine}
\usepackage{bbding}
\usepackage{xspace}
\usepackage{textcomp}
\usepackage{textcase}
\usepackage{setspace}

\newcommand {\B}{\textcolor {blue}}

\def\para{\ensuremath{/\kern -0.8em /}\xspace}

\def\beqn{\begin{eqnarray}}
\def\eeqn{\end{eqnarray}}
\def\beq{\begin{equation}}
\def\eeq{\end{equation}}

\newcommand{\Beq}{\begin{eqnarray*} }
\newcommand{\Eeq}{\end{eqnarray*} }
\newcommand{\Bmat}{\left(\begin{matrix}}
\newcommand{\Emat}{\end{matrix}\right)}

\graphicspath{{./}}

\begin{document}

\title{Kitaev-derived Gapless Spin Liquid in the $K$-$J$-$\Gamma$-$\Gamma'$ Quantum Magnet Na$_2$Co$_2$TeO$_6$}

\author{Han Li}
\email{hanli@bjtu.edu.cn}
\affiliation{School of Physical Science and Engineering, Beijing Jiaotong University, Beijing 100044, China}

\author{Xu-Guang Zhou}
\affiliation{Anhui Key Laboratory of Low-Energy Quantum Materials and Devices, High Magnetic Field Laboratory (CHMFL), Hefei Institutes of Physical Science, Chinese Academy of Sciences, Hefei, 230031, China}

\author{Gang Su}
\affiliation{Institute of Theoretical Physics, Chinese Academy of Sciences, Beijing 100190, China}
\affiliation{Kavli Institute for Theoretical Sciences, University of Chinese Academy of Sciences, Beijing 100190, China}

\author{Wei Li}
\email{w.li@itp.ac.cn}
\affiliation{Institute of Theoretical Physics, Chinese Academy of Sciences, Beijing 100190, China}

\begin{abstract} 
\end{abstract}
\date{\today}
\maketitle

\noindent
\textbf{The realization of quantum spin liquids (QSLs) in Kitaev magnets represents an intriguing topic in frustrated quantum magnetism. Despite prediction in the pure Kitaev honeycomb model, realization of QSLs in realistic systems and materials remain scarce. The recent discovery of cobalt-based compound Na$_2$Co$_2$TeO$_6$ has raised significant research interest. By establishing a realistic $K$-$J$-$\Gamma$-$\Gamma'$ model for Na$_2$Co$_2$TeO$_6$ --- with a dominant antiferromagnetic (AFM) Kitaev interaction ($K>0$) that quantitatively explains its thermodynamics measurements --- we reveal an intermediate gapless QSL phase under [111] magnetic fields with tensor-network calculations. We confirm the QSL nature of this phase by demonstrating its adiabatic connection to the intensively studied intermediate QSL of the pure AFM Kitaev model under out-of-plane fields. Our results show excellent agreement with recent high-field experiments, thereby explaining the intermediate-field phase in Na$_2$Co$_2$TeO$_6$. These findings bridge the gap between theoretical proposals for a Kitaev-derived QSL and experimental realization, opening new avenues for exploring exotic quantum states of matter in realistic Kitaev materials.
}
\\

\noindent{\bf{Introduction}}\\
Quantum spin liquids (QSLs), an exotic class of topological state of matter beyond the traditional symmetry-breaking paradigm, has garnered significant research interest~\cite{Anderson1973, Balents2010, Zhou2017, JW2019QMats, Broholm2020Science}. Amongst others, the Kitaev honeycomb model, with bond-directional Ising-type couplings, is exactly soluble and features the Kitaev QSL~\cite{Kitaev2003, Kitaev2006, Hermanns2018}. Under a finite field, there emerges non-Abelian anyon in the Kitaev model, a cornerstone for topological quantum computations~\cite{Kitaev2003, Kitaev2006}. In light of the Jackeli-Khaliullin mechanism~\cite{Jackeli2009}, substantial progresses have been made in the material realization of the Kitaev QSL, especially for the prime candidate material $\alpha$-RuCl$_3$~\cite{Winter2016, Winter2017NC, Wu2018, Cookmeyer2018, Kim2016, Suzuki2019, Ran2017, Wang2017, Ozel2019, Banerjee2016, HSKim2015}.
Experimental studies reveal that $\alpha$-RuCl$_3$ constitutes a proximate Kitaev QSL material with strong ferromagnetic (FM) Kitaev interactions, with clear signatures of spin fractionalization at finite temperature or frequency~\cite{Do2017, Banerjee2018, Li2020b, Han2021, Han2023}.

Besides the $4d$-magnet $\alpha$-RuCl$_3$, there are other Kitaev candidate materials emerging recently. Theoretical studies suggest that the peculiar Kitaev interactions 
may also appear in certain $3d$ honeycomb magnets. 
Owing to the high-spin $d^7$ electronic configuration of Co$^{2+}$ and the edge-sharing CoO$_6$ octahedra, it gives rise to a pseudo-spin $J_{\rm eff} = \frac{1}{2}$ ground state and potentially Kitaev interactions between the Co$^{2+}$ ions~\cite{Liu2020PRL}. Thus the Co-based compounds have attracted significant research interest~\cite{Lin2021NC, Yao2022PRL, Li2022, Zhong2020SA, Zhang2023NatMat, Halloran2023PNAS, Yao2023}. Among them, Na$_2$Co$_2$TeO$_6$ (NCTO) features nearly ideal honeycomb layers and exhibits antiferromagnetic (AFM) long-range order below $T_{\rm N} \approx 27$~K~\cite{Lefrancois2016PRB, Bera2017PRB}.
The nature of the AFM order, either a zigzag magnetic structure~\cite{Lefrancois2016PRB, Bera2017PRB, Zhang2023PRB, Samarakoon2021PRB} or a triple-Q ground state configuration~\cite{Chen2021PRB, Lee2021}, remains actively debated. Under an in-plane magnetic field along $a^*$ axis, a first-order phase transition occurs near 6~T~\cite{Yao2020} or 7.5~T~\cite{Lin2021NC}. When the field is greater than 9.5 T, it enters the (partially) polarized phase~\cite{Yao2020}. However, the AFM order is much harder to be suppressed under the out-of-plane fields along the $c$ axis. Such a strong magnetic anisotropy cannot be simply ascribed to the Land\'e factor, reminiscent of similar phenomena reported in $\alpha$-RuCl$_3$~\cite{Lampen-Kelley2018, Banerjee2017, Weber2016}.

Although significant effort has been devoted to elucidate the magnetic properties of NCTO and its microscopic spin model~\cite{Liu2020PRL,Lin2021NC,Yao2022PRL,Li2022,Zhong2020SA,Zhang2023NatMat, Halloran2023PNAS, Lefrancois2016PRB, Bera2017PRB, Zhang2023PRB, Chen2021PRB, Lee2021, Yao2020}, the identification of Kitaev interactions in this material remains an active area of investigation. Prior efforts have suggested both the $K$-$J_{1,2,3}$-$\Gamma$-$\Gamma'$ and XXZ $J_1$-$J_3$ models as candidate models for NCTO~\cite{Songvilay2020, Kim2021, Lin2021NC, LinG2024, Samarakoon2021PRB, Li2022}, however, even the sign of the Kitaev interactions remains unsettled. Some studies propose FM Kitaev coupling in NCTO based on linear spin wave theory~\cite{Songvilay2020} , while others argue for AFM interactions based on inelastic neutron scattering and related analyses~\cite{Kim2021, Lin2021NC, LinG2024}. 
Thus, determining the microscopic spin model for NCTO 
constitutes the first step in understanding its magnetic properties,
which requires advanced many-body thermodynamic calculations.

In this work, we first systematically investigate the ground-state and thermodynamic properties of these candidate models. Our analysis reveals that all of these models exhibit significant difficulties in reproducing the experimental specific heat results using the thermal tensor network approach~\cite{LTRG, Chen2018, Li2022tan}. Therefore, through careful fittings of the specific heat profiles and spin structure factors, we propose an alternative effective $K$-$J$-$\Gamma$-$\Gamma'$ model with dominant AFM Kitaev interactions $K \approx 19$~meV for NCTO, which can well explain the key experimental observations with a single set of model parameters. These prominent experimental observations include: (i) double-peak specific heat with peaks corresponding to temperatures 
of magnitude $O(10)$~K and $O(100)$~K, (ii) strong magnetic anisotropy (in-plane vs. out-of-plane), and (iii) field-induced quantum phase transitions, along different directions (in-plane $a^*$ and out-of-plane $c$ axes). We also obtain the temperature-field phase diagram, from which we uncover two distinct quantum phase transitions under out-of-plane magnetic fields, evidenced by spin structure factors, isentropic curves, and Gr\"uneisen parameter results, etc. We thus identify the intermediate-field phase as a gapless QSL, possibly quantum Majorana metal, which is adiabatically connected to that of the pure AFM Kitaev model under [111] (out-of-plane) field. Our work firmly establishes NCTO as a viable platform for realizing field-induced QSL states, and pave the theoretical ground for its further experimental studies.
\\

\noindent{\bf{Results}}\\
\textbf{Effective $K$-$J$-$\Gamma$-$\Gamma'$ model with dominant AFM Kitaev term.} Two different types of effective spin models have been proposed to describe the magnetic properties of NCTO~\cite{Songvilay2020, Kim2021, Lin2021NC, LinG2024, Samarakoon2021PRB, Kim2021, Li2022}, which we have conducted a systematic evaluation (see Supplementary Note.~\B{1}). Although most of the proposed models can reproduce the experimentally observed zigzag magnetic order~\cite{Lefrancois2016PRB, Bera2017PRB, Zhang2023PRB, Samarakoon2021PRB} via density matrix renormalization group (DMRG) calculations, the comparison between our finite-temperature tensor-network calculations (see Methods) and experimental measurements reveals a clear inconsistency: we find absence of double-peak structre in the simulated specific heat of XXZ $J_1$-$J_3$ model; while certain $K$-$J$-$\Gamma$-$\Gamma'$ models possess double-peak specific heat~\cite{Lefrancois2016PRB, Yao2020, Li2022}, the peak positions are found different from experiments (see {Supplementary Note.~\B{1}}). This inconsistency motivates us to find an optimized parameter set that simultaneously stabilizes the zigzag-ordered ground state and quantitatively reproduces the finite-temperature behaviors observed in experiments.

The effective $K$-$J$-$\Gamma$-$\Gamma'$ model for NCTO reads:
\begin{equation}
\begin{split}
H=& \sum_{\langle i,j\rangle_{\gamma}} [K S_i^{\gamma}S_j^{\gamma} + J\,\textbf{S}_i\cdot \textbf{S}_j 
+\Gamma(S_i^{\alpha}S_j^{\beta}+S_i^{\beta}S_j^{\alpha}) \\
& +\Gamma'(S_i^{\gamma}S_j^{\alpha}+S_i^{\gamma}S_j^{\beta}+S_i^{\alpha}S_j^{\gamma}+S_i^{\beta}S_j^{\gamma})],
\end{split}
\label{Eq:HamRuCl3}
\end{equation}
where $\langle i,j\rangle_{\gamma}$ indicates the nearest neighbor sites connected with the $\gamma$-type bond, $\textbf{S}_i=\{S^x_i, S^y_i, S^z_i\}$ represents the spin-$1/2$ operators on the $i$-th site, and $\{ \alpha, \beta, \gamma \}$ refers to a cyclic permutation of $\{ x, y, z \}$; $K$ is the Kitaev interaction; $J$ is the nearest-neighbor isotropic Heisenberg interaction; $\Gamma$ and $\Gamma'$ are the off-diagonal couplings. 

Inspired by the striking similarity in magnetic and thermodynamics behaviors between the two compounds NCTO and $\alpha$-RuCl$_3$~\cite{Kubota2015, Widmann2019, Sears2015, Banerjee2017, Do2017}, we take such $\alpha$-RuCl$_3$ model~\cite{Han2021} as the starting point for constructing the effective model for NCTO. In Ref.~\cite{Han2021}, the model parameters for $\alpha$-RuCl$_3$ determined from the thermal data are $\{K, J, \Gamma, \Gamma'\}/|K|$ = $\{-1, -0.1, 0.3, -0.02\}$ with $|K| = 25$~meV. On the other hand, previous studies points to an AFM Kitaev term ($K>0$) in NCTO~\cite{Kim2021, Lin2021NC, LinG2024}. Nevertheless, we notice that the effective models of two Kitaev candidate materials can be related via a unitary transformation~\cite{Chaloupka2015}. Through a global $\pi$ rotation around $c$ axis, we can transform the $\alpha$-RuCl$_3$ model, with the Kitaev coupling becoming positive, and then fine tune it to fit the thermodynamic measurements (see Methods). We finally arrive at a parameter set $\{K, J, \Gamma, \Gamma'\}/K$ = $\{1, -0.9035, -0.65, 0.3597\}$ with $K = 19$~meV for NCTO. The field orientation $H_{[l m n]}$ is defined in the spin $\{S^x, S^y, S^z\}$ axes, where $H_{[1 1 1]}$ represents the out-of-plane $c$-direction field. The Zeeman term is thus $g_{[l m n]} \mu_{\rm B} \mu_{0} H_{[l m n]} \frac{l S^x + m S^y + n S^z} {\sqrt{l^2+m^2+n^2}}$, where $g_{[1 1 1]} =g_{\rm c} \simeq 2.3$ and $g_{[\bar{1} 0 1]} = g_{\rm ab} \simeq 4.33$~\cite{Lin2021NC}. With this set of proposed parameters, we show below that the model calculations can well reproduce the thermodynamics measurements, including the specific heat with two characteristic peaks occur at temperatures differing by almost an order of magnitude~\cite{Yao2022PRL,Yang2022}, magnetic anisotropy along in-plane and out-of-plane directions~\cite{Yao2020, Lin2021NC, Zhang2023PRB, Xiao2022}, and finite-temperature evolution of spin-structure factors~\cite{Li2022, LinG2024, Yao2023}, etc.
\\

\begin{figure}[t!]
\includegraphics[angle=0,width=1\linewidth]{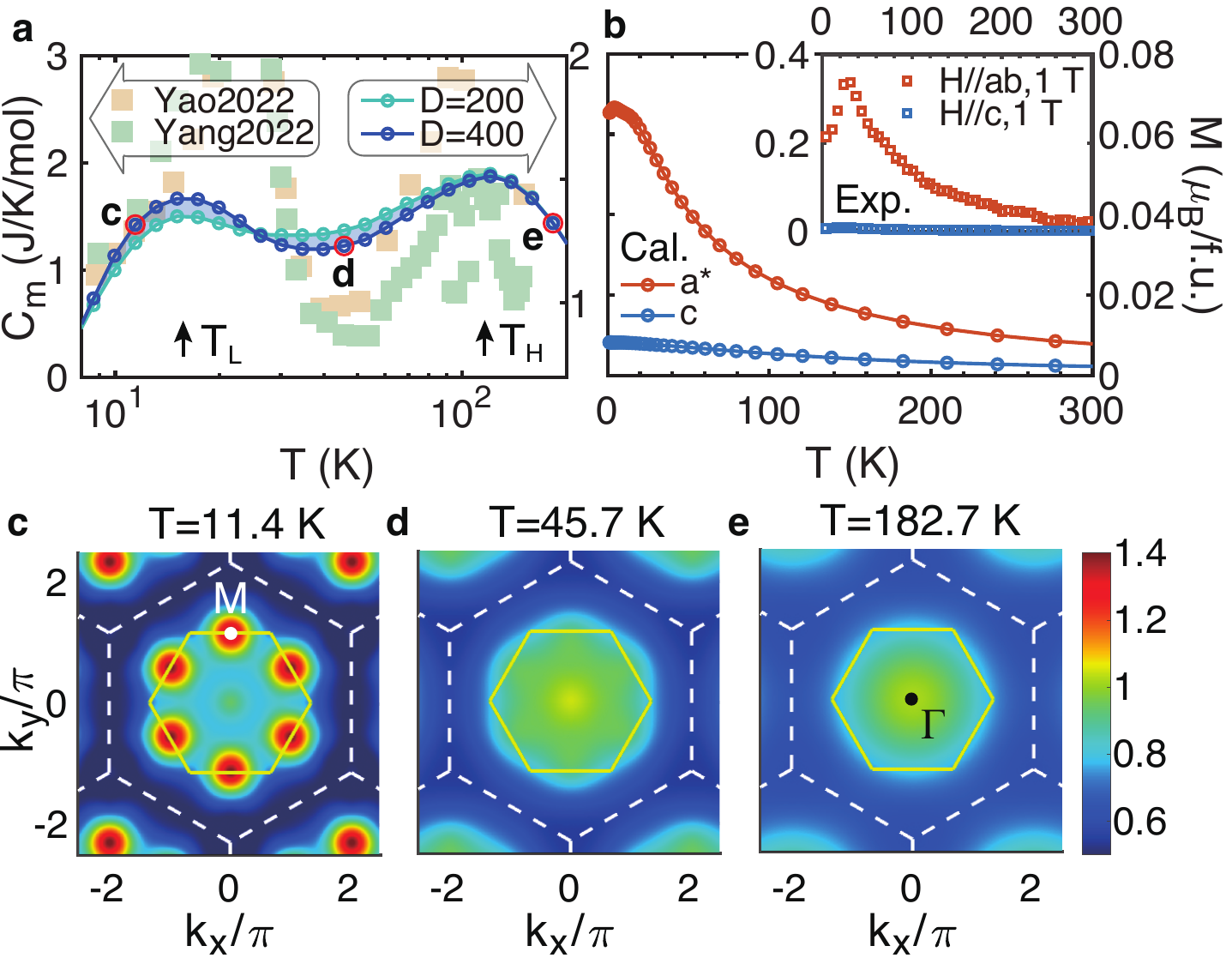}
\renewcommand{\figurename}{\textbf{Fig.}}
\caption{\textbf{Comparison between model simulations and experimental results.} 
\textbf{a} Zero-field specific heat $C_{\rm m}$ curves versus temperature. Experimental data (squares, left axis) with legends ``Yao2022'' and ``Yang2022'' from Refs.~\cite{Yao2022PRL,Yang2022} exhibit characteristic double peaks at temperature scales $T_{\rm L}$ and $T_{\rm H}$. Theoretical calculations (blue curves with open circles, right axis) display systematic convergence with increasing bond dimension ($D=200$ and $400$).
\textbf{b} shows the calculated results of in-plane and out-of-plane magnetization $M(T)$ at 1 Tesla, with experimental data~\cite{Xiao2022} shown in the inset. Both theoretical and experimental data exhibit strong magnetic anisotropy.
\textbf{c-e} The spin structure factors $S(\textbf{q})$ (see the main text) at:
\textbf{c} low-, \textbf{d} intermediate-, and \textbf{e} high-temperature regimes, with their temperature values indicated in panel \textbf{a}. The results are symmetrized over the equivalent ${\textbf q}$ points.
}
\label{Fig:OurPara}
\end{figure}

\begin{figure*}[t!]
\includegraphics[angle=0,width=1\linewidth]{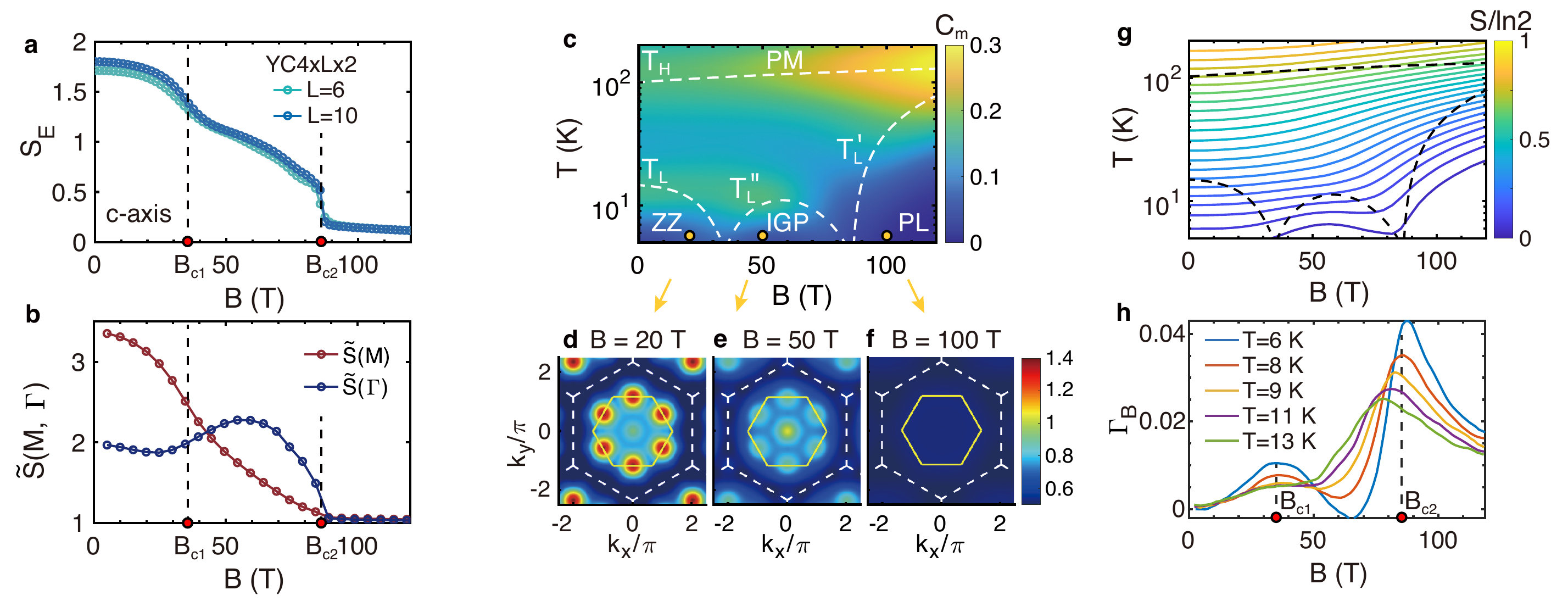}
\renewcommand{\figurename}{\textbf{Fig.}}
\caption{\textbf{Field-induced phase transitions in NCTO under out-of-plane fields.} 
\textbf{a} Zero-temperature entanglement entropy $S_{\rm E}$ with different system sizes. Two transition fields $B_{\rm c1} \simeq 35$~T and $B_{\rm c2} \simeq 86$~T are detected and labeled by the red dots.
\textbf{b} The spin structure factors ${\tilde S}(\textbf{q})$ (see main text) curves at $q=M,\Gamma$ points, from which the transition fields can also be determined.
\textbf{c} The landscape of specific heat $C_{\rm m}$. The light color highlights the $C_{\rm m}$ maxima, and the white dashed lines indicate the schematic phase boundaries separating paramagnetic (PM), zigzag (ZZ), intermediate gapless phase (IGP), and polarized (PL) phases.
\textbf{d-f} The spin structure factor results ${\tilde S}(\textbf{q})$ under fields $B=20$, $50$ and $100$~T at $T=5.7$~K, which reflect the characteristics of three distinct low-temperature field-induced phases ZZ, IGP and PL, respectively.
\textbf{g} The isentropic curves starting from various temperature, and the color code corresponds to the values of normalized thermal entropy $S/{\rm ln}2$. 
\textbf{h} The Gr\"uneisen parameter $\Gamma_{\rm B}$ results at various low temperatures. The dashed lines mark fields $B_{\rm c1}$ and $B_{\rm c2}$, corresponding to the two $\Gamma_{\rm B}$ peaks, which indicate two distinct quantum phase transitions in the system.
}
\label{Fig:PD111}
\end{figure*}

\noindent
\textbf{Magnetic properties and thermodynamics.}
In Fig.~\ref{Fig:OurPara}\textbf{a}, by performing thermal tensor renormalization group (XTRG~\cite{Chen2018}) simulations based on the NCTO model parameters as mentioned above, we present the specific heat $C_{\rm m}$ curves with two different bond dimensions $D$, which are well converged with each other and lead to reliable results. The experimental data are shown in the background with solid squares, which are adapted from Refs.~\cite{Yao2022PRL,Yang2022} from two different research groups. As the experimental magnetic specific heat $C_{\rm m}$ peaks exhibit slight variations in their absolute magnitudes amongst different experiments, we emphasize the importance of the characteristic double-peak structure and their respective temperature scales $T_{\rm H}$ and $T_{\rm L}$, which provide key information for determination of model parameters. Our parameter set could well reproduce $T_{\rm H}$ and $T_{\rm L}$ at respectively 100~K and around 20~K.
The small difference around low-temperature peak position $T_{\rm L}$ likely arises from finite-size effects inherent to the cylindrical geometry in the simulations.

With the same set of model parameters, we calculate the magnetization curves along $a^*$ and $c$ axes, evaluated at $1$~T. As shown in Fig.~\ref{Fig:OurPara}\textbf{b}, the in-plane magnetization is nearly one order of magnitude larger than the out-of-plane value, indicating that the system exhibits strong magnetic anisotropy. It is consistent with the experimental results~\cite{Xiao2022} as shown in the inset. Consequently, we find our model is capable of reproducing both the specific heat characteristics and the magnetic anisotropy along different field directions.

At finite temperatures and zero magnetic field, we show in Fig.~\ref{Fig:OurPara}\textbf{c-e} the static spin structure factors ${S}(\textbf{q})=\sum_{j} e^{i \textbf{q} (\textbf{r}_j-\textbf{r}_{i_0})} \langle S_{i_0} S_j\rangle$, where the sub-index $i_0$ represents a central site of the lattice, and the summation is taken over all lattice sites. At $T=11.4$~K in Fig.~\ref{Fig:OurPara}\textbf{c}, we observe large intensities at M points in the Brillouin zone (BZ), which is a signature of the low-temperature AFM order~\cite{Li2022, LinG2024, Yao2023}. We note that it has been ascribed to a zigzag~\cite{Lefrancois2016PRB, Bera2017PRB, Zhang2023PRB} or a triple-Q~\cite{Chen2021PRB, Lee2021,Hong2024} order in experiments, while our DMRG results here favor the zigzag order (see Supplementary Note.~\B{1}). Moreover, the $\Gamma$-point intensity is less pronounced compared to that observed in $\alpha$-RuCl$_3$ results~\cite{Han2021}, which may be attributed to the relatively stronger non-Kitaev $|\Gamma|$ interaction present in this system than in the $\alpha$-RuCl$_3$ magnet. At intermediate temperature $T=45.7$~K, the renowned M-star structure present in $\alpha$-RuCl$_3$~\cite{Banerjee2017} also appears here in NCTO model calculations shown in Fig.~\ref{Fig:OurPara}\textbf{d}, which is indicative of the spin fractionalization due to the Kitaev coupling. At high temperature, the system enters a conventional paramagnetic regime, evidenced by the broad and diffusive peak at the $\Gamma$ point in Fig.~\ref{Fig:OurPara}\textbf{e}.
\\

\noindent
\textbf{Two quantum phase transitions induced by out-of-plane fields.}
Through XTRG and ground-state DMRG calculations, we map out the temperature-field phase diagram along two different field directions. Under in-plane fields along the $a^*$ axis, we find a single phase transition between the zigzag AFM and polarized phases (see {Supplementary Note.~\B{2}}); while under out-of-plane $c$-direction fields, as shown in Fig.~\ref{Fig:PD111}, two transition fields are observed at $B_{\rm c1} \simeq 35$~T and $B_{\rm c2} \simeq 86$~T, respectively. It is noteworthy that $B_{\rm c1}$ and $B_{\rm c2}$ obtained from our model calculations are consistent with $H_{\rm c}^{\rm AFM}$ and $H_{\rm c}^{\rm S}$ observed in the recent high-field magnetization experiments~\cite{Zhou2024arXiv}.

The ground-state bipartite entanglement entropy $S_{\rm E}$ results can be used to determine the transition fields, as presented in Fig.~\ref{Fig:PD111}\textbf{a}. At around $B_{\rm c1} \simeq 35$~T and $B_{\rm c2}\simeq 86$~T, the $S_{\rm E}$ curve exhibits anomalies. The determined critical fields are in excellent agreement with the transition fields identified in the experimental measurements. The experimental out-of-plane magnetic susceptibility exhibits two phase transitions $H_{\rm c}^{\rm AFM}$ and $H_{\rm c}^{\rm S}$ for small field angles to the $c^*$-axis~\cite{Zhou2024arXiv}. 

The transition fields $B_{\rm c1}$ and $B_{\rm c2}$ can also be determined through the spin-structure factors ${\tilde S}(\textbf{q})$ with background subtracted, i.e., $$\tilde S (\textbf{q})=\sum_{j} e^{ i \textbf{q} (\textbf{r}_j-\textbf{r}_{i_0})} (\langle S_{i_0} S_j\rangle - \langle S_{i_0} \rangle \langle S_j \rangle).$$ 
As shown in Fig.~\ref{Fig:PD111}\textbf{b}, at small fields, the system exhibits a zigzag AFM phase at low temperatures, as indicated by the structure factor at M point (see the red curve in Fig.~\ref{Fig:PD111}\textbf{b}). For higher fields, as the system enters the spin polarized phase, both $\tilde S (M)$ and $\tilde S (\Gamma)$ values vanish. 
Therefore, the positions where $\tilde S (M)$ decreases rapidly, and where both $\tilde S (M)$ and $\tilde S (\Gamma)$ values approach vanish, mark the two quantum phase transitions (see also in Figs.~\ref{Fig:IntermediateField2}\textbf{a,b}).
Under intermediate fields, the $\tilde S (M)$ is decreased, with its magnitude even below that of $\tilde S (\Gamma)$, indicating the suppression of long-range order. Such characteristics can also be observed in Figs.~\ref{Fig:PD111}\textbf{d-f} at a low temperature $T=5.7$~K, under fields $B=20$, 50 and 100~T, respectively. In the following section, we will discuss the nature of this intermediate-field phase.

\begin{figure}[h!]
\includegraphics[angle=0,width=0.99\linewidth]{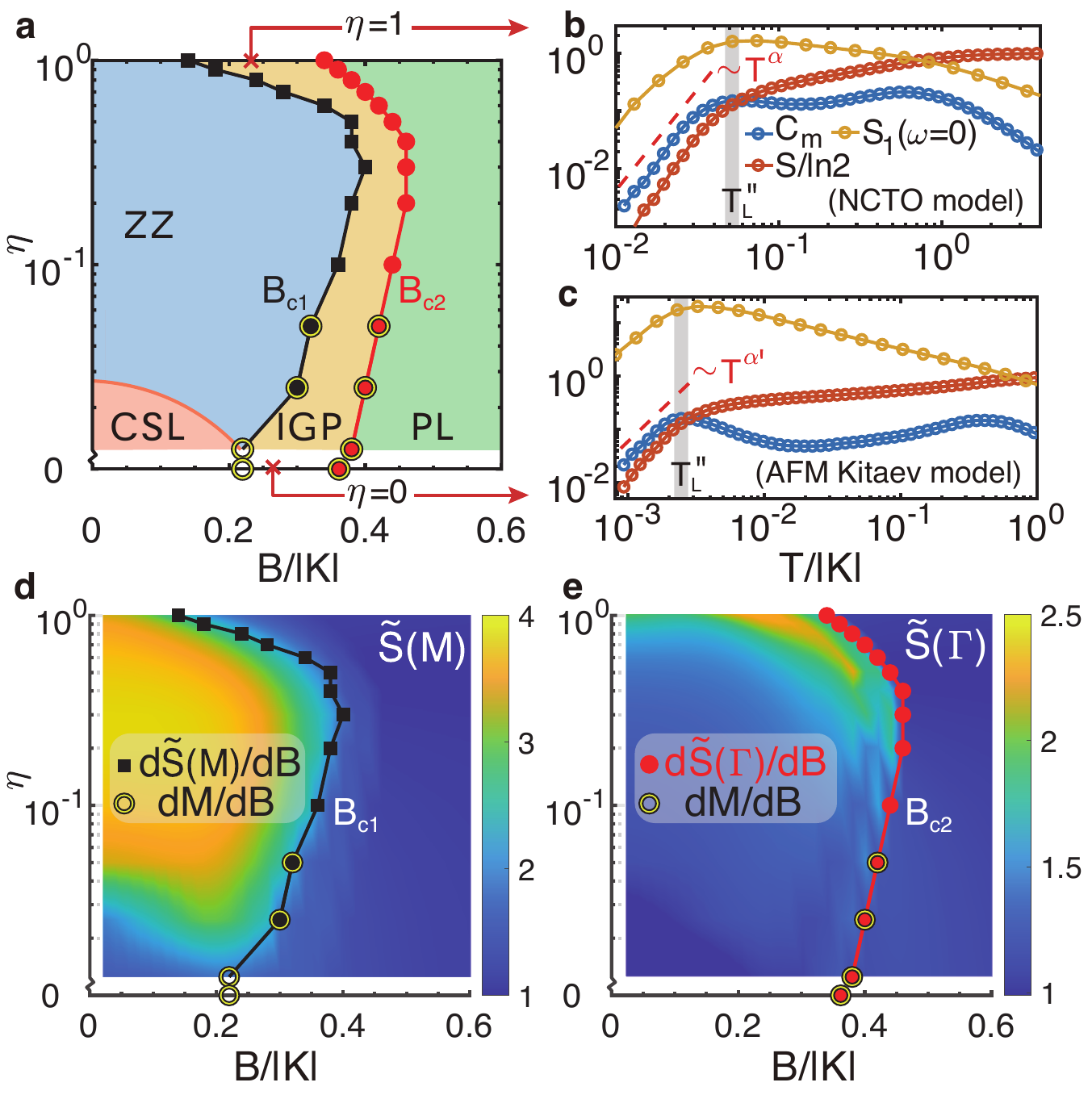}
\renewcommand{\figurename}{\textbf{Fig.}}
\caption{\textbf{$\eta$-$B$ phase diagram connecting pure AFM Kitaev and NCTO model.} 
\textbf{a} The $\eta$-$B$ phase diagram constructed from structure factor results, where CSL, ZZ, IGP and PL denote chiral spin liquid, zigzag order, intermediate-field gapless phase, and spin polarized phase, respectively.
\textbf{b,c} Low-temperature thermodynamic results including the specific heat $C_{\rm m}$, thermal entropy $S/{\rm ln}2$, and the spin-lattice relaxation rate $S_1(\omega=0)$ (see main text), at characteristic fields $B/|K| \simeq 0.26$ and 0.23 within the IGP for $\eta=0$ and 1, respectively 
(see the red crosses marked in \textbf{a}). The gray vertical bar marks the characteristic temperature scale $T_{\rm L}''$, below which we observed a power-law scaling as indicated by the red dashed line. 
\textbf{d,e} Contour plots of spin structure factors ${\tilde S}(M)$ and ${\tilde S}(\Gamma)$ under different fields $B/|K|$ and parameters $\eta$ (see main text) values. The transition fields $B_{\rm c1}$ and $B_{\rm c2}$ are determined from ${\rm d}{\tilde S}(q=M,\Gamma)/{\rm d}B$ and ${\rm d}M/{\rm d}B$ curves (see Fig.~\ref{Fig:IntermediateField2} and Supplementary Note.~\B{3}).
The results are presented in natural unit in these plots.
}
\label{Fig:IntermediateField}
\end{figure}

\begin{figure}[h!]
\includegraphics[angle=0,width=0.97\linewidth]{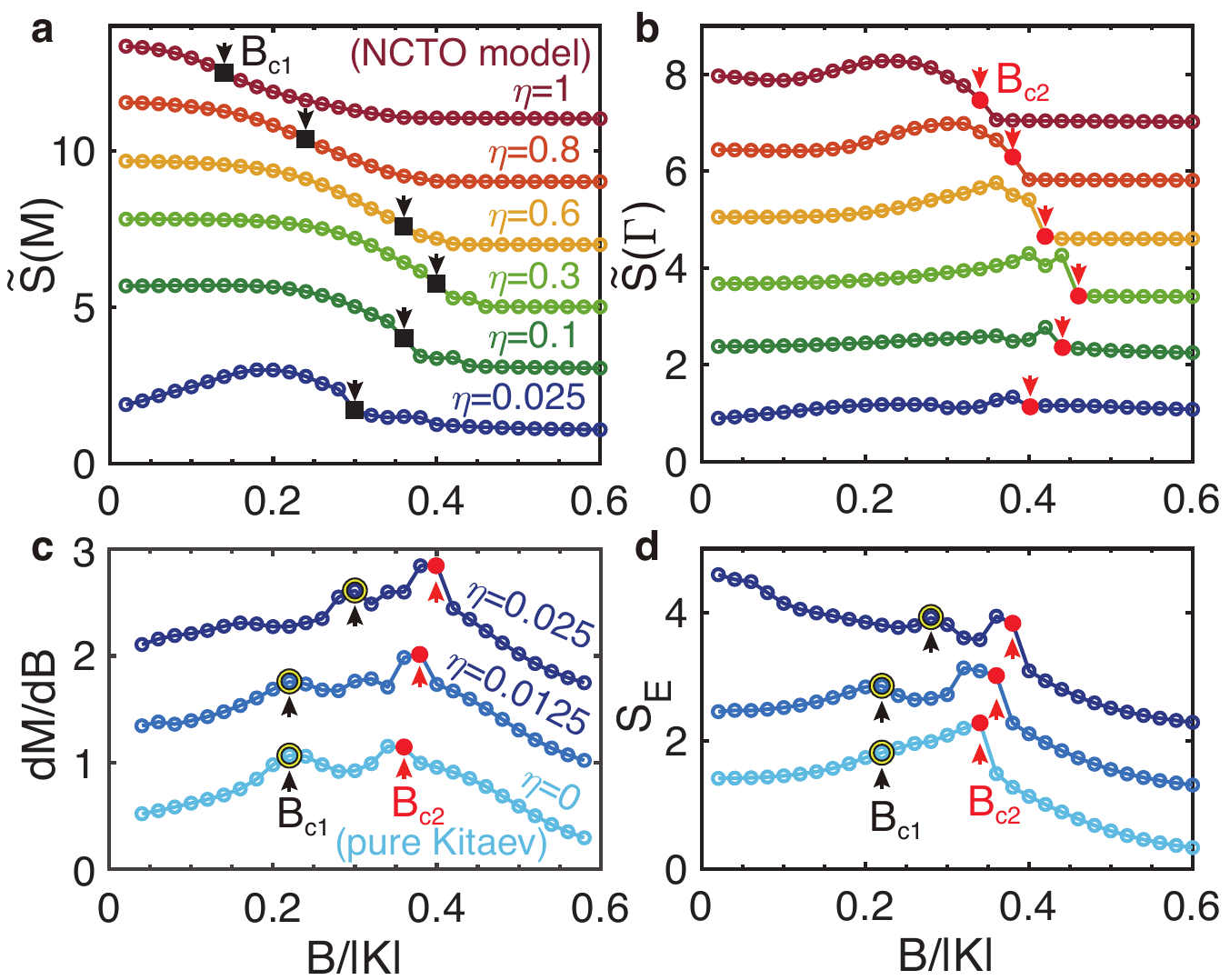}
\renewcommand{\figurename}{\textbf{Fig.}}
\caption{{\textbf{Determination of the phase boundary.} 
\textbf{a,b} The ${\tilde S}(M)$ and ${\tilde S}(\Gamma)$ curves as a function of magnetic field $B/|K|$ for different $\eta \geq 0.025$. The black and red arrows mark the dips of ${\rm d}{\tilde S}(q=M,\Gamma)/{\rm d}B$ curves at transition fields $B_{\rm c1}$ and $B_{\rm c2}$, respectively. 
\textbf{c} The derivatives of magnetization ${\rm d}M/{\rm d}B$ and \textbf{d} the bipartite entanglement entropy $S_{\rm E}$ curves at various small $\eta$, where the peaks indicate $B_{\rm c1}$ and $B_{\rm c2}$ and are indicated by the black and red arrows.
{The curves in these plots have been vertically offset for clarity.}}
}
\label{Fig:IntermediateField2}
\end{figure}

For finite-temperature calculations, we present the landscape of the specific heat $C_{\rm m}$ in Fig.~\ref{Fig:PD111}\textbf{c}, and mark out the temperature scales $T_{\rm H}$, $T_{\rm L}$, $T_{\rm L}'$, and $T_{\rm L}''$. Upon increasing magnetic fields, the high-temperature scale $T_{\rm H}$ firstly demonstrates a relatively flat line, persistent across both the low-field zigzag-ordered phase and the intermediate-field phase at low temperatures. When the field is further enhanced, $T_{\rm H}$ gradually shifts towards higher temperatures and finally merge with the Zeeman energy scale $T_{\rm L}'$, producing a broad maximum in $C_{\rm m}$ as visualized in the intensity plot (Fig.~\ref{Fig:PD111}\textbf{c}, upper right). In the intermediate-field regime, a lower-temperature hump/peak feature emerges in $C_{\rm m}$ at $T_{\rm L}''$, separated from $T_{\rm H}$ (also see Fig.~\ref{Fig:IntermediateField}\textbf{c}), which corresponds to the dome-like feature in Fig.~\ref{Fig:PD111}\textbf{g} and will be discussed below. Therefore, we find the three low-temperature phases, i.e., low-field zigzag ordered AFM phase, intermediate-field disordered phase, and high-field polarized phase, are separated by $B_{\rm c1}$ and $B_{\rm c2}$ as indicated in Fig.~\ref{Fig:PD111}\textbf{c}.


The low-temperature isentropic curves of thermal entropy $S/{\rm ln}2$ can also determine the two phase transitions. In Fig.~\ref{Fig:PD111}\textbf{g}, it shows an increasement near 35~T upon entering the intermediate-field regime (as indicated by the black dashed dome-like boundary), accompanied by pronounced valley features upon leaving this regime. Derived from the simulated isentrope data, in Fig.~\ref{Fig:PD111}\textbf{h} we show the calculated Gr\"uneisen parameter at low temperatures, i.e., $\Gamma_{\rm B} = \frac{1}{T}(\frac{\partial T}{\partial B})_S$. 
Near $B_{\rm c1}$ there is a single peak without sign reversal, while near $B_{\rm c2}$ a peak-dip structure with clear sign reversal is observed. It suggests that there is a first- ($B_{\rm c1}$) and second-order phase transitions ($B_{\rm c2}$) at the two transition fields.
\\

\noindent
\textbf{Gaplss QSL under out-of-plane fields.}
To explore the nature of IGP obtained from the realistic NCTO model, we construct an $\eta$-$B$ phase diagram connecting $K$-$J$-$\Gamma$-$\Gamma'$ and the pure AFM Kitaev honeycomb model. The existence of an intermediate QSL phase under out-of-plane magnetic fields has been intensively studied, yet its fundamental nature remains unresolved~\cite{Gohlke2018, Zhu2018, Liang2018, Jiang2018, Trebst2019, Patel2019, Jiang2020, Jahromi2021, Han2024}. Recent work suggests this state is a quantum Majorana metal, characterized by itinerant Majorana fermions moving through a background of fluctuating $Z_2$ fluxes~\cite{Zhu2025QMM}. In Fig.~\ref{Fig:IntermediateField}, we introduce a tuning parameter $\eta$ to modulate the strength of non-Kitaev interactions including the $J$, $\Gamma$ and $\Gamma'$ terms, i.e., $H = H_{\rm Kitaev} + \eta \, H'_{\rm NCTO}$, with 
\begin{equation}
\begin{split}
H'_{\rm NCTO}=& \sum_{\langle i,j\rangle_{\gamma}} [ J\,\textbf{S}_i\cdot \textbf{S}_j + \Gamma(S_i^{\alpha}S_j^{\beta}+S_i^{\beta}S_j^{\alpha}) \\
& +\Gamma'(S_i^{\gamma}S_j^{\alpha}+S_i^{\gamma}S_j^{\beta}+S_i^{\alpha}S_j^{\gamma}+S_i^{\beta}S_j^{\gamma}) ], 
\end{split}
\label{Eq:HamAFKitaev2NCTO}
\end{equation}
where $\gamma \in \{x,y,z\} $ and $\eta = 0$ and $1$ represent the pure AFM Kitaev model and NCTO model, respectively. 

We summarize our results in the $\eta$-$B$ phase diagram in Fig.~\ref{Fig:IntermediateField}\textbf{a}, where the phase boundaries $B_{\rm c1}$ and $B_{\rm c2}$ separate ZZ, IGP, and PL phases for large $\eta$ values, while CSL, IGP, and PL phases for very small $\eta$ values. These phase boundaries are determined from the spin structure factors $\tilde S (\textbf{q})$ as detailed in Figs.~\ref{Fig:IntermediateField}\textbf{d,e} and Fig.~\ref{Fig:IntermediateField2}. The gapless nature of the IGP is supported by the simulated thermodynamic properties in Fig.~\ref{Fig:IntermediateField}\textbf{b}. Under a typical field strength $B/|K| \simeq 0.23$ within the IGP of NCTO system, which corresponds to $B\simeq 57$~T in experiments, we find there exist algebraic scaling $\sim T^{\alpha}$ below the low-temperature scale $T^{''}_{\rm L}$ in specific heat $C_{\rm m}$, thermal entropy $S/{\rm ln}2$ and spin-lattice relaxation rate estimated by $S_1(\omega=0) = \frac{1}{T}\sum_{\gamma}\sum_{j=1}^{N}[\langle S_j^{\gamma}(\frac{\beta}{2}) S_j^{\gamma}(0)\rangle - \langle S_j^{\gamma}(\beta) \rangle^2]$~\cite{Xi2023}. More data supporting the gapless nature can be found in the {Supplementary Note.~\B{3}}. Therefore, the IGP exhibits neither magnetic ordering nor a gap, which supports an out-of-plane field-induced gapless QSL state in NCTO magnet.

It is particularly noteworthy that the IGP in NCTO exhibits common properties to the intermediate-field QSL of the pure AFM Kitaev model at $\eta \simeq 0$. In Fig.~\ref{Fig:IntermediateField}\textbf{c}, we selected an intermediate-field strength $B/|K| \simeq 0.26$ and computed the same thermodynamic quantities as in Fig.~\ref{Fig:IntermediateField}\textbf{b}, observing a similar algebraic low-temperature scaling $\sim T^{\alpha'}$. These results indicate that the IGP in the realistic NCTO system may directly connect to the gapless QSL of the ideal AFM Kitaev model. Below, we demonstrate the connectivity of the two phases through tuning the parameter $\eta$ and field strength $B$ based on zero-temperature DMRG calculations.

The critical fields can be determined from the spin structure factors $\tilde S (\textbf{q})$. In Fig.~\ref{Fig:IntermediateField}\textbf{d} and Fig.~\ref{Fig:IntermediateField}\textbf{e}, we present contour plots of the M- and $\Gamma$-point spin structure factors, and the phase boundaries are clearly visualized. At larger values of $\eta$, the system remains in the ZZ phase under small fields, leading to a pronounced M-point structure factor $\tilde S(M)$ as shown in Fig.~\ref{Fig:IntermediateField}\textbf{d}. As the field increases into the IGP, the AFM order is suppressed, leading to a rapid decrease in $\tilde S(M)$. The lower critical field, $B_{\rm c1}$, is therefore identified from this drop in $\tilde S(M)$, with the extracted values marked by black squares (see also in Fig.~\ref{Fig:IntermediateField2}\textbf{a}). For reference, the derivative ${\rm d}\tilde S(M)/{\rm d}B$ is provided in Supplementary Note.~\B{3}.

The high-field phase transition, in contrast, is characterized by the $\Gamma$-point structure factors $\tilde S(\Gamma)$. By subtracting the static contribution from field polarization, $\tilde S(\textbf{q})$ more clearly reveals the intrinsic correlations in the system. In the high-field PL phase, $\tilde S(\textbf{q})$ vanishes, as the intensity there in the spin structure factor is entirely due to field polarization (Fig.~\ref{Fig:IntermediateField}\textbf{e}). The transition field $B_{\rm c2}$ to the PL phase is thus identified by the point where $\tilde S(\Gamma)$ drops to zero, as marked with red circles in Fig.~\ref{Fig:IntermediateField2}\textbf{b}.

In the small-$\eta$ regime, the system resides in a CSL phase under low magnetic fields, we use different criterion to determine the phase boundary between the CSL and IGP. In Fig.~\ref{Fig:IntermediateField2}\textbf{c}, we show the derivative of the magnetization ${\rm d}M/{\rm d}B$ and identify the positions of two transition fields. We benchmark our method using $\eta = 0.025$, where the lower transition field (from the ZZ to the IGP phase) is consistent whether determined from $\tilde{S}(M)$ or ${\rm d}M/{\rm d}B$. For smaller $\eta$ in the CSL phase, the positions of transition fields are determined from ${\rm d}M/{\rm d}B$, which are found to be consistent with those determined from bipartite entanglement entropy $S_{\rm E}$ results (see Fig.~\ref{Fig:IntermediateField2}\textbf{d}). Near the two transition fields, the $S_{\rm E}$ curves exhibits peaks or humps. Based on these results, especially the phase diagram in Fig.~\ref{Fig:IntermediateField}\textbf{a}, we conclude that the IGP of NCTO model and that in pure AFM Kitaev model are adiabatically connected.
\\

\noindent{\bf{Discussion}} \\
The present work establishes a $K$-$J$-$\Gamma$-$\Gamma'$ model for the Kitaev material NCTO. Through systematic evaluation of previously proposed model parameters for NCTO~\cite{Songvilay2020, Kim2021, Lin2021NC, LinG2024, Samarakoon2021PRB, Li2022}, combining XTRG ($T>0$) and DMRG ($T=0$) calculations, we demonstrate that while most 
models correctly reproduce the zigzag-ordered ground state, they exhibit significant discrepancies with experimental specific heat data. It thus motivates us to find an optimized parameter set that account both the ground-state and finite-temperature properties. Inspired by the striking similarities in magnetic properties between the compound NCTO and the prime Kitaev magnet $\alpha$-RuCl$_3$, we propose a model that demonstrates quantitative agreement with recent experimental observations on NCTO~\cite{Yao2022PRL, Yang2022, Yao2020, Lin2021NC, Zhang2023PRB, Xiao2022, Li2022, LinG2024, Yao2023, Zhou2024arXiv}, including:
(i) the characteristic double-peak structure in specific heat at temperatures of magnitude $O(10)$~K and $O(100)$~K,
(ii) strong magnetic anisotropy along $a^*$ and $c$ axes,
(iii) spin-structure factors across various temperatures, 
and (iv) high-field magnetization curves along in- and out-of-plane directions.

More importantly, our model predicts that the Kitaev candidate NCTO hosts a gapless QSL in the intermediate-field regime. This phase can be adiabatically connected to the intermediate-field phase of the pure AFM Kitaev model. While the specific nature of this QSL in the pure model --- whether a gapped $\mathbb{Z}_2$ state~\cite{Jiang2020, Zhang2022}, a gapless U(1) spinon Fermi surface~\cite{Jiang2018, Patel2019, Han2024}, or a gapless $\mathbb{Z}_2$ quantum Majorana metal~\cite{Wang2025PRB, Zhu2025QMM} --- remains under active investigation, the presence of a QSL is well-established. Therefore, our calculations predict that the intermediate-field phase observed in recent experimental on NCTO~\cite{Zhou2024arXiv} is a manifestation of this Kitaev-derived QSL. Thus, although realistic materials inevitably contain non-Kitaev interactions that obscure the realization of the pure Kitaev model, we propose that an out-of-plane magnetic field can suppress these terms, allowing a quantum spin liquid primarily ascribed to the pure Kitaev model to emerge.

Notably, the existence of the IGP phase has recently been experimentally identified in NCTO~\cite{Zhou2024arXiv}. We further propose future experimental protocols for studying this gapless QSL state at high fields ($>35$~T):
(i) low-temperature thermodynamic measurements such as the specific heat that can reveal the gapless behavior below 10~K;
(ii) magnetic torque measurements, which should exhibit a very small total net torque value in the low-field zigzag AFM phase, and then increase rapidly in the intermediate QSL regime~\cite{Han2023};
(iii) NMR measurements at intermediate fields can reveal a power-law dependence $1/T_1 \sim T^{\alpha}$, with the exponent $\alpha$ identifying the gapless nature of the intermediate-field QSL phase. Lastly, we propose that this scenario of Kitaev-derived QSL apply to not only NCTO but also $\alpha$-RuCl$_3$: the intermediate-field phase in $\alpha$-RuCl$_3$~\cite{Han2021, Zhou2023} --- characterized by the suppression of zigzag order and the emergence of QSL phase --- may also be adiabatically connects to this Kitaev-derived QSL. We expect this scenario can guide future theoretical and experimental studies on Kitaev materials, including NCTO and $\alpha$-RuCl$_3$, as well as other Kitaev candidates like BaCo$_2$(AsO$_4$)$_2$~\cite{Zhong2020SA, Halloran2023PNAS, Maksimov2025PRL}. 
\\

\noindent{\bf{Methods}}\\
\textbf{Zero and finite-temperature tensor network approaches.}
The ground-state and finite-temperature properties are calculated 
by the density matrix renormalization group (DMRG) method~\cite{WhitePRL},
recognized as an effective approach for calculating two-dimensional (2D) lattice systems,
and the exponential tensor renormalization group (XTRG) method~\cite{Chen2018},
which has been previously applied to study several 2D finite-size spin systems~\cite{Chen2018,
Chen2018b,Lih2019,Li2020b,Li2020,Hu2020,Han2021,Gao2022,Han2023,
Wang2023PRL, Xiang2024, Han2024}, respectively. 
In particular, for studying the Kitaev system 
featuring two temperature scales where one of which is remarkably low,
XTRG method has shown its accuracy and high efficiency~\cite{Li2020b, Han2021, Han2023, Han2024}.
The system sizes we adopt in the ground-state and the finite-temperature simulations in this work
are cylindrical lattices with width $W=4$, length $L=6$ or $10$,
the total number of lattice sites $N=W\times L\times 2$ and periodic boundary conditions along the $W$ direction (Y-type cylinder, YC). The bond dimension $D$ is up to $512$ with truncation error $\epsilon \lesssim 1\times 10^{-7}$ in DMRG, and $400$ with $\epsilon$ less than $1\times 10^{-3}$ in XTRG down to $T/|K| \simeq 0.01$. The convergency check of different $D$ and different sizes $L$ are shown in Fig.~\ref{Fig:OurPara}\textbf{a} and Fig.~\ref{Fig:PD111}\textbf{d}, which gives well-converged results.\\

\noindent
\textbf{Unitary transformation and the coupling parameters.} 
For the $K$-$J$-$\Gamma$-$\Gamma'$ model, the unitary transformation is
$$
\begin{pmatrix}
J \\
K \\
\Gamma \\
\Gamma'
\end{pmatrix}^{'}=\tau
\begin{pmatrix}
J \\
K \\
\Gamma \\
\Gamma'
\end{pmatrix}=
\begin{pmatrix}
1  & +\frac{4}{9} & -\frac{4}{9} &+\frac{4}{9} \\
0 & -\frac{1}{3}  & +\frac{4}{3} & -\frac{4}{3}\\
0 & +\frac{4}{9} & +\frac{5}{9} & +\frac{4}{9}\\
0 & -\frac{2}{9} & +\frac{2}{9} & +\frac{7}{9}
\end{pmatrix}
\begin{pmatrix}
J \\
K \\
\Gamma \\
\Gamma'
\end{pmatrix},
$$
where $\tau$ is a global $\pi$ rotation about the $[1 1 1]$ axis (perpendicular to the honeycomb plane)~\cite{Chaloupka2015}. Therefore, the proposed parameters for $\alpha$-RuCl$_3$, i.e., 
$\{K, J, \Gamma, \Gamma'\}/|K|$ = $\{-1, -0.1, 0.3, -0.02\}$ with $|K| = 25$~meV~\cite{Han2021}
could be transformed to $\{K, J, \Gamma, \Gamma'\}/|K|$ = $\{1, -0.9035, -0.3772, 0.3597\}$ with $|K| = 19$~meV. Through tuning the $\Gamma/|K|$ ratio to $-0.65$, we could simultaneously reproduce both the specific heat and field-induced transitions in NCTO. The details of the parameter fitting process could be found in {Supplementary Note.~\B{1}}.\\

\noindent{\bf{Data availability}}\\
Source data are provided in this paper. 
The data generated in this study have been deposited in the Zenodo database.\\

\noindent{\bf{Code availability}}\\
All numerical codes in this paper are available 
upon request to the authors. \\

\bibliography{kitaevRef}

$\,$\\
\textbf{Acknowledgements} \\
The authors acknowledge supports by the National Natural Science Foundation of China (Grants Nos. 12404177 (H.L.), 12222412 and 12047503 (W.L.)), the Strategic Priority Research Program of Chinese Academy of Sciences, Grant No. XDB1270100 (W.L.), and the Talent Fund of Beijing Jiaotong University (Grant No. 2025JBRC003) (H.L.). We thank HPC-ITP for the technical support and generous allocation of CPU time.

$\,$\\
\textbf{Author contributions} \\
H.L. and W.L. initiated this work. 
H.L. performed the tensor-network calculations.
All the authors conducted theoretical analysis. 
H.L. and W.L. prepared the manuscript with input from all authors.
All authors reviewed the manuscript.

$\,$\\
\textbf{Competing interests} \\
The authors declare no competing interests. 

$\,$\\
\textbf{Additional information} \\
\textbf{Supplementary Information} is available in the online version of the paper. \\
\noindent

\clearpage
\newpage
\onecolumngrid
\begin{center}
{\large Supplementary Information for}
$\,$\\ 
$\,$\\ 
\textbf{\large{Kitaev-derived Gapless Spin Liquid in the $K$-$J$-$\Gamma$-$\Gamma'$ Quantum Magnet Na$_2$Co$_2$TeO$_6$}}
$\,$\\
$\,$\\ 
Li \textit{et al.}
\end{center}

\newpage


\date{\today}

\setcounter{subsection}{0}
\setcounter{figure}{0}
\setcounter{equation}{0}
\setcounter{table}{0}

\renewcommand{\thesubsection}{\normalsize{Supplementary Note \arabic{subsection}}}
\renewcommand{\theequation}{S\arabic{equation}}
\renewcommand{\thefigure}{\arabic{figure}}
\renewcommand{\thetable}{\arabic{table}}

\subsection{Model parameters for Na$_2$Co$_2$TeO$_6$}
\label{SecSM:PreviousModels}

\textbf{Summary of previously proposed models for Na$_2$Co$_2$TeO$_6$.}
For the Kitaev magnet Na$_2$Co$_2$TeO$_6$ (NCTO) material, 
a series of model parameters have been previously proposed in the literature~\cite{Songvilay2020, Kim2021, Lin2021NC, Samarakoon2021PRB, LinG2024, Li2022}. 
Here we systematically summarize and organize these parameters 
in Supplementary Table.~\ref{Tab:Parameters1} and Supplementary Table.~\ref{Tab:Parameters2}. 
These model parameters can be classified into two descriptions, 
which are respectively described by 
$K$-$J_{1,2,3}$-$\Gamma$-$\Gamma'$ and XXZ $J_1$-$J_3$ frameworks. 
For the $K$-$J_{1,2,3}$-$\Gamma$-$\Gamma'$ model, 
we have normalized all parameters with respect to the Kitaev interaction $|K|$, 
while for XXZ $J_1$-$J_3$ model, 
the normalization was performed relative to the $J_3^{xy}$ interaction. 
The corresponding energy scales (in meV units) 
are all derived from their respective references.

\begin{table*}[h]
\begin{tabular}{llllllllllll}
\hline
\multicolumn{1}{|c|}{Refs.} 
& \multicolumn{1}{c|}{\ $K$ (meV)\ } & \multicolumn{1}{c|}{\ $\Gamma/|K|$ \ } & \multicolumn{1}{c|}{\ $\Gamma'/|K|$\ } & \multicolumn{1}{c|}{\ $J_1/|K|$\ } & \multicolumn{1}{c|}{\ $J_2/|K|$\ } & \multicolumn{1}{c|}{\ $J_3/|K|$\ } & \multicolumn{1}{c|}{\ $K_{\rm fit}$ (meV)\ } & \multicolumn{1}{c|}{\ zigzag\ } & \multicolumn{1}{c|}{\ $C_{\rm m}$\ } \\ \hline

\multicolumn{1}{|c|}{Songvilay2020~\cite{Songvilay2020}} & \multicolumn{1}{c|}{-9}  & \multicolumn{1}{c|}{0.2}  & \multicolumn{1}{c|}{0.033} &  \multicolumn{1}{c|}{-0.011
} & \multicolumn{1}{c|}{0.033}& \multicolumn{1}{c|}{0.1} & \multicolumn{1}{c|}{-28} & \multicolumn{1}{c|}{\Checkmark} & \multicolumn{1}{c|}{\Checkmark}\\

\multicolumn{1}{|c|}{Kim2021~\cite{Kim2021}} & \multicolumn{1}{c|}{3.3}  & \multicolumn{1}{c|}{-0.848}  & \multicolumn{1}{c|}{0.636} &  \multicolumn{1}{c|}{-0.4545} & \multicolumn{1}{c|}{0}& \multicolumn{1}{c|}{0.4545} & \multicolumn{1}{c|}{10} & \multicolumn{1}{c|}{\Checkmark} & \multicolumn{1}{c|}{\Checkmark}\\

\multicolumn{1}{|c|}{Lin2021~\cite{Lin2021NC}} & \multicolumn{1}{c|}{0.125}  & \multicolumn{1}{c|}{1}  & \multicolumn{1}{c|}{0} &  \multicolumn{1}{c|}{-18.6} & \multicolumn{1}{c|}{0}& \multicolumn{1}{c|}{20} & \multicolumn{1}{c|}{0.25} & \multicolumn{1}{c|}{\Checkmark} & \multicolumn{1}{c|}{\XSolidBrush}\\

\multicolumn{1}{|c|}{Samarakoon2021A~\cite{Samarakoon2021PRB}} & \multicolumn{1}{c|}{-7}  & \multicolumn{1}{c|}{0.00286}  & \multicolumn{1}{c|}{-0.03286} &  \multicolumn{1}{c|}{-0.0286} & \multicolumn{1}{c|}{0.0071}& \multicolumn{1}{c|}{0.1714} & \multicolumn{1}{c|}{-28} & \multicolumn{1}{c|}{\Checkmark} & \multicolumn{1}{c|}{\Checkmark}\\

\multicolumn{1}{|c|}{Samarakoon2021B~\cite{Samarakoon2021PRB}} & \multicolumn{1}{c|}{2.7}  & \multicolumn{1}{c|}{-1.074}  & \multicolumn{1}{c|}{0.5926} &  \multicolumn{1}{c|}{-1.1852} & \multicolumn{1}{c|}{0.037}& \multicolumn{1}{c|}{0.4444} & \multicolumn{1}{c|}{10} & \multicolumn{1}{c|}{\Checkmark} & \multicolumn{1}{c|}{\Checkmark} \\

\multicolumn{1}{|c|}{Lin2022~\cite{LinG2024}} & \multicolumn{1}{c|}{1.28}  & \multicolumn{1}{c|}{-0.9375}  & \multicolumn{1}{c|}{0.625} &  \multicolumn{1}{c|}{-1.0938} & \multicolumn{1}{c|}{0}& \multicolumn{1}{c|}{0.9375} & \multicolumn{1}{c|}{3.8} & \multicolumn{1}{c|}{\Checkmark} & \multicolumn{1}{c|}{\XSolidBrush}\\

\multicolumn{1}{|c|}{our model} & \multicolumn{1}{c|}{19}  & \multicolumn{1}{c|}{-0.9035}  & \multicolumn{1}{c|}{-0.65} &  \multicolumn{1}{c|}{0.3597} & \multicolumn{1}{c|}{0}& \multicolumn{1}{c|}{0} & \multicolumn{1}{c|}{19} & \multicolumn{1}{c|}{\Checkmark} & \multicolumn{1}{c|}{\Checkmark} \\
\hline
\end{tabular}

\renewcommand{\tablename}{\textbf{Supplementary Table }}
\caption{\textbf{Parameters and checklist of various effective $K$-$J_{(1,2,3)}$-$\Gamma(')$ 
model proposed for Na$_2$Co$_2$TeO$_6$.}
The ``\Checkmark'' and ``\XSolidBrush'' symbols
are used to indicate whether the experimentally observed characteristics 
can be reproduced by the corresponding parameter set.
In the table, ``zigzag'' means a formation of ground-state zigzag AFM order
(see Supplementary Fig.~\ref{FigS1:SqMz}),
while ``$C_{\rm m}$'' means the double-peak specific heat feature 
(see Supplementary Fig.~\ref{FigS:Cm}).
}
\label{Tab:Parameters1}
\end{table*}

\begin{table*}[h]
\begin{tabular}{llllllllllll}
\hline
\multicolumn{1}{|c|}{Refs.} 
& \multicolumn{1}{c|}{\ $J_3^{xy}$ (meV)\ } & \multicolumn{1}{c|}{\ $J_3^z/|J_3^{xy}|$ \ } & \multicolumn{1}{c|}{\ $J_1^{xy}/|J_3^{xy}|$\ } & \multicolumn{1}{c|}{\ $J_1^z/|J_3^{xy}|$\ } & \multicolumn{1}{c|}{\ $D/|J_3^{xy}|$\ } & \multicolumn{1}{c|}{\ $J_{3,\rm fit}^{xy}$ (meV)\ } & \multicolumn{1}{c|}{\ zigzag\ } & \multicolumn{1}{c|}{\ $C_{\rm m}$\ }\\ \hline

\multicolumn{1}{|c|}{Kim2021B~\cite{Kim2021}} & \multicolumn{1}{c|}{2.1}  & \multicolumn{1}{c|}{0.95}  & \multicolumn{1}{c|}{-1} &  \multicolumn{1}{c|}{-0.95} & \multicolumn{1}{c|}{0.1506}& \multicolumn{1}{c|}{5} & \multicolumn{1}{c|}{\Checkmark} & \multicolumn{1}{c|}{\XSolidBrush} \\\hline

\multicolumn{1}{|c|}{Yao2022~\cite{Yao2022PRL}} & \multicolumn{1}{c|}{1.896}  & \multicolumn{1}{c|}{1}  & \multicolumn{1}{c|}{0} &  \multicolumn{1}{c|}{0} & \multicolumn{1}{c|}{0.2175}& \multicolumn{1}{c|}{5} & \multicolumn{1}{c|}{\XSolidBrush} & \multicolumn{1}{c|}{\XSolidBrush} \\\hline

\end{tabular}

\renewcommand{\tablename}{\textbf{Supplementary Table }}
\caption{\textbf{Parameters and checklist of various effective XXZ-$J_1$-$J_3$ models proposed for Na$_2$Co$_2$TeO$_6$.}
The ``\Checkmark'' and ``\XSolidBrush'' symbols
are used to indicate whether the experimentally observed characteristics 
can be reproduced by the corresponding parameter set.
In the table, ``zigzag'' means a formation of ground-state zigzag AFM order
(see Supplementary Fig.~\ref{FigS1:SqMz}),
while ``$C_{\rm m}$'' means the double-peak specific heat feature 
(see Supplementary Fig.~\ref{FigS:Cm}).
}
\label{Tab:Parameters2}
\end{table*}

\textbf{Ground-state spin correlations.}
Employing these model parameters, 
we applied the density matrix renormalization group (DMRG) method 
to calculate the ground-state spin correlations and magnetic moments 
for each parameter set listed in Supplementary Table.~\ref{Tab:Parameters1} and Supplementary Table.~\ref{Tab:Parameters2}. 
The calculations were performed on YC$8\times 12\times 2$ lattice sites
(e.g., see Supplementary Fig.~\ref{FigS1:SqMz}\textbf{a'}), 
with a kept bond dimension $D=512$. 
Supplementary Fig.~\ref{FigS1:SqMz}\textbf{a-h} and \textbf{a'-h'} 
show the corresponding spin structure factors $S(\textbf{q})$
and the expectation values of magnetic moments $\langle S^{\gamma}\rangle$ ($\gamma = x,y,z$). 
Through the spin structure factors results in Supplementary Fig.~\ref{FigS1:SqMz}\textbf{a-h}, 
we observe that most parameter sets (with the exception of the ``Yao2022'' set~\cite{Yao2022PRL}
as shown in Supplementary Fig.~\ref{FigS1:SqMz}\textbf{h}) 
exhibit prominent intensity at the M-point in the Brillouin zone (BZ). 
Combined with their magnetic configuration results on the right-hand side, 
this indicates the formation of a zigzag antiferromagnetic (AFM) magnetic order in the ground state,
consistent with the experimental observations~\cite{Li2022, LinG2024, Yao2023}. 
Accordingly, we have added a column started with ``zigzag'' in Supplementary Table.~\ref{Tab:Parameters1} and Supplementary Table.~\ref{Tab:Parameters2}, 
where checkmarks (``\Checkmark'') and crosses (``\XSolidBrush'') 
are used to denote whether the experimentally observed zigzag AFM order 
can be reproduced by each parameter set.

\begin{figure*}[h!]
\includegraphics[angle=0,width=0.9\linewidth]{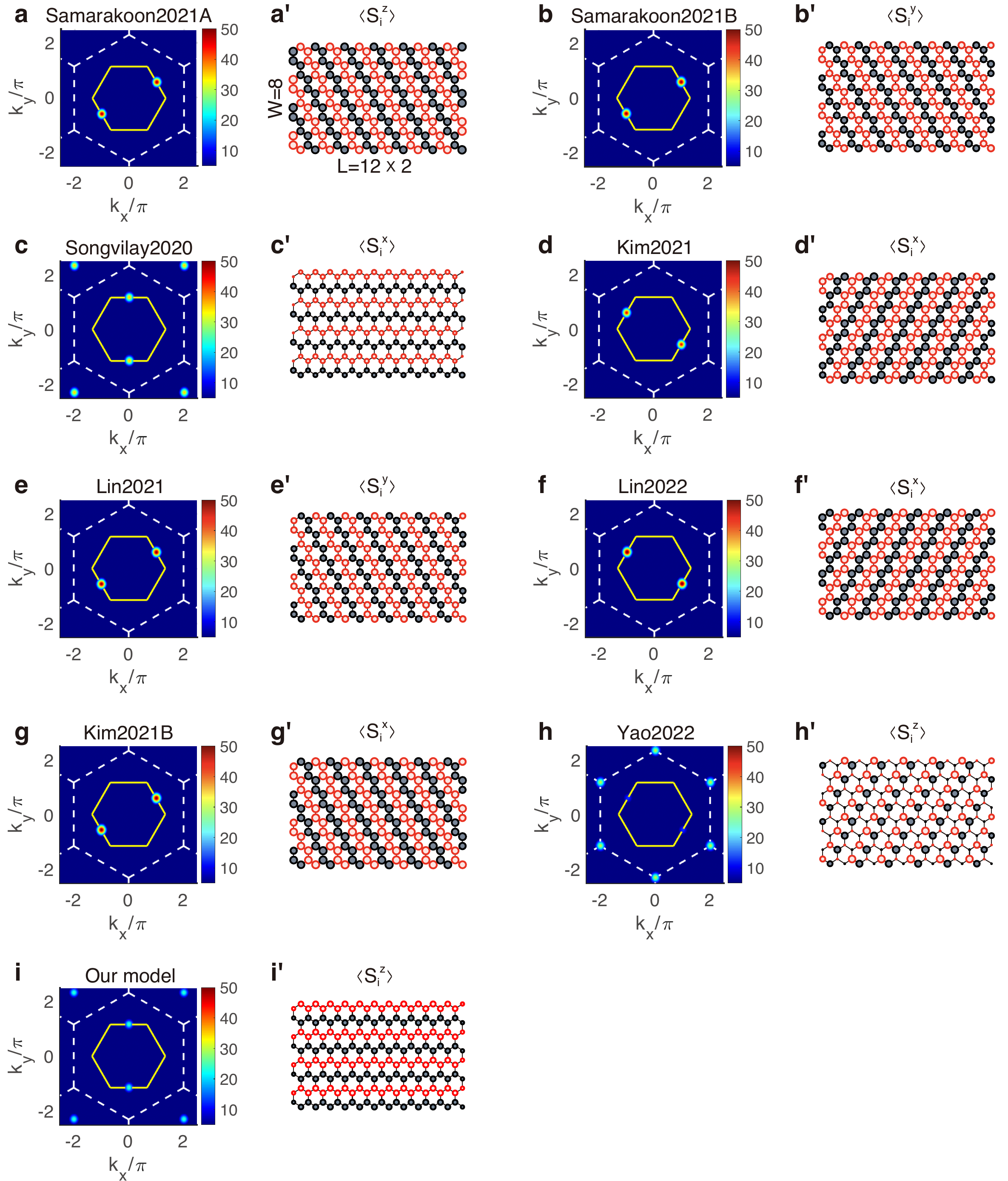}
\renewcommand{\figurename}{\textbf{Supplementary Figure }}
\caption{\textbf{Ground-state spin correlations of various proposed models.} 
\textbf{a-i} The spin structure factors $S(\textbf{q})$ (see main text for details) 
under different parameter sets with a unified color code applied across all panels. 
\textbf{a'-i'} The corresponding on-site magnetic moment configurations $\langle S^{\gamma} \rangle$  
along a specific component $\gamma=x,y$ or $z$, 
where red circles denote positive and black circles represent negative expectation values. 
The circle size corresponds to the magnitude of $\langle S^{\gamma} \rangle$ values.
}
\label{FigS1:SqMz}
\end{figure*}

\textbf{Finite-temperature specific heat results.}
In addition to the ground-state spin correlations, 
we further calculated the finite-temperature specific heat $C_{\rm m}$ curves
for each parameter set using the exponential tensor renormalization group (XTRG) method~\cite{Chen2018}, 
as presented in Supplementary Fig.~\ref{FigS:Cm}. 
The arrangement of panels in Supplementary Fig.~\ref{FigS:Cm} maintains 
a one-to-one correspondence with panels \textbf{a-h} in Supplementary Fig.~\ref{FigS1:SqMz}. 
The gray curves with open circles represent calculated results 
obtained directly from the proposed parameters $K$ or $J_{\rm 3}^{xy}$~\cite{Songvilay2020, Kim2021, Lin2021NC, Samarakoon2021PRB, LinG2024, Li2022}, 
while light-colored square symbols denote experimental data 
from two independent studies~\cite{Yao2022PRL,Yang2022}. 
We find that all parameter sets yield specific heat curves 
showing significant deviations from experimental measurements, 
thereby obscuring the thermal characteristics 
within the temperature range of interest.
We therefore rescaled the temperature axes for each parameters in each panels
by only adjusting the energy scale from 
the original $K$ or $J_{\rm 3}^{xy}$ values in Supplementary Table.~\ref{Tab:Parameters1} and Supplementary Table.~\ref{Tab:Parameters2} 
to optimized $K_{\rm fit}$ and $J_{\rm 3, fit}^{xy}$ values, 
with the rescaled results shown as blue curves with open circles. 
Considering that the experimental specific heat 
exhibits a characteristic double-peak feature, 
our systematic comparison reveals that only parameter sets 
corresponding to panels \textbf{a-d} successfully reproduce this distinctive structure
(although it does not means a specific heat matching with experiments, 
as the energy scale is manual shifted). 
Consequently, we have added a ``$C_{\rm m}$'' column in Supplementary Table.~\ref{Tab:Parameters1} and Supplementary Table.~\ref{Tab:Parameters2}, 
where (``\Checkmark'') and crosses (``\XSolidBrush'') indicate 
the presence or absence of the experimentally observed 
double-peak specific heat signature for each parameter set.

\begin{figure*}[h]
\includegraphics[angle=0,width=1\linewidth]{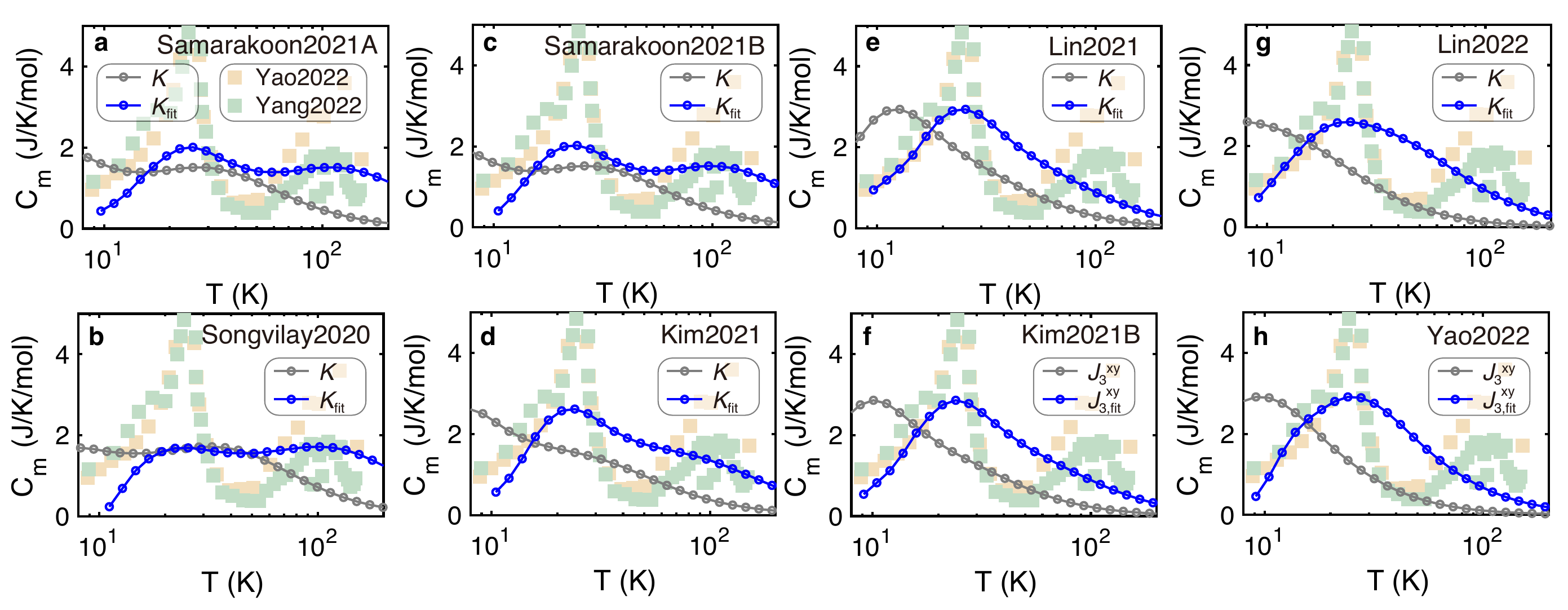}
\renewcommand{\figurename}{\textbf{Supplementary Figure }}
\caption{\textbf{Comparison of specific heat $C_{\rm m}$ curves calculated from 
various proposed models with experimental data.} 
The grey curves with open circles represent the results calculated strictly 
using the proposed energy scale $K$ or $J_{\rm 3}^{xy}$, 
while the blue ones depict the corresponding results with
adjusted energy scale $K_{\rm fit}$ and $J_{\rm 3, fit}^{xy}$ for each case.
The solid squares denote the experimental data~\cite{Yao2022PRL,Yang2022}. 
In panels \textbf{a-d}, the magnetic specific heat $C_{\rm m}$
displays a double-peak structure when plotted with the optimized energy scale $K_{\rm fit}$,
where the peak positions show agreement with experimental observations. 
In contrast, panels \textbf{e-h} reveal that $C_{\rm m}$
exhibits only a single peak structure. 
The specific values of the optimized parameters $K_{\rm fit}$ and $J_{\rm 3, fit}^{xy}$
are provided in Supplementary Table.~\ref{Tab:Parameters1} and Supplementary Table.~\ref{Tab:Parameters2}.
}
\label{FigS:Cm}
\end{figure*}

\textbf{Our parameter set.}
Although the four aforementioned parameter sets 
presented in Supplementary Fig.~\ref{FigS:Cm}\textbf{a-d} successfully 
reproduce the double-peak specific heat characteristics, 
we observe that the necessary energy-scale modifications 
(via $K \to K_{\rm fit}$) required to fit the specific heat curves 
inevitably lead to discrepancies between the calculated zero-temperature transition fields 
and experimental observations~\cite{Zhou2024arXiv}.
Therefore, we aim to propose a revised parameter set 
that could simultaneously satisfy both thermodynamic (specific heat peak positions) 
and field-induced properties (in-plane and out-of-plane transition fields) constraints.

\begin{figure*}[h]
\includegraphics[angle=0,width=0.55\linewidth]{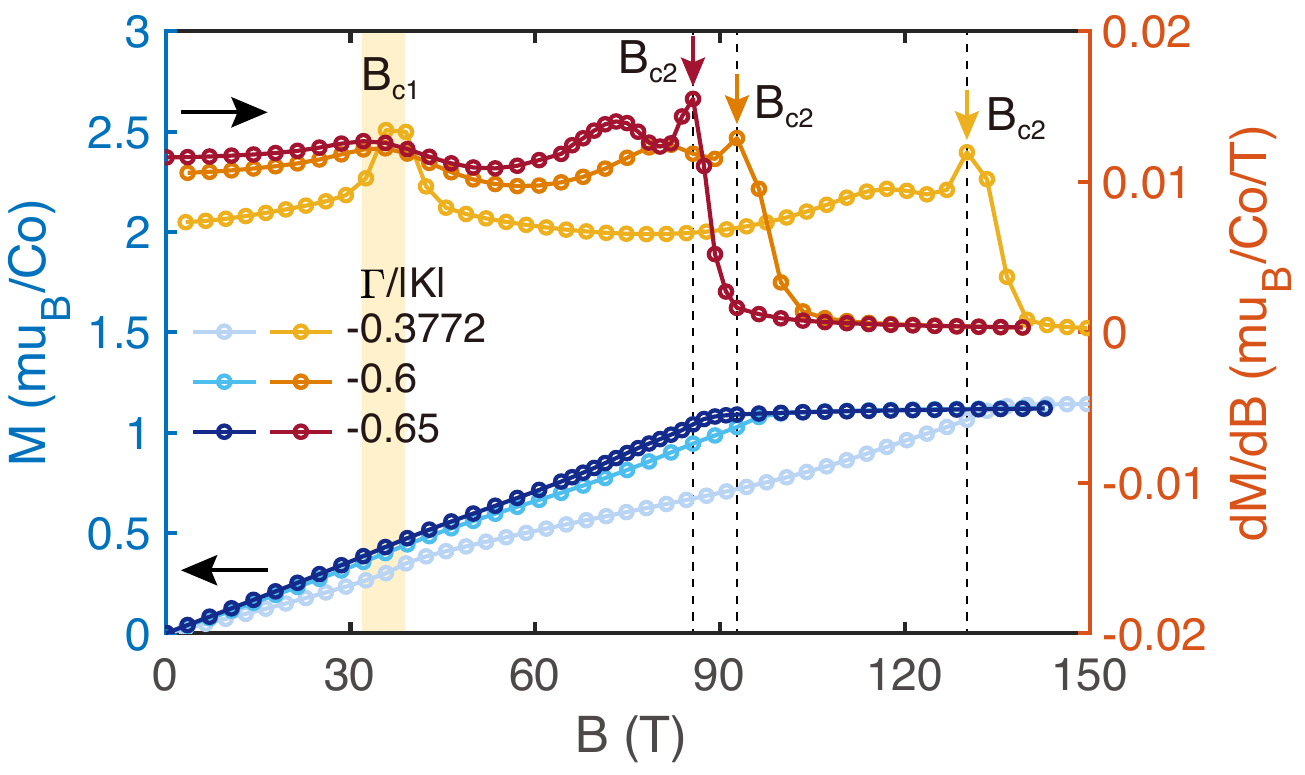}
\renewcommand{\figurename}{\textbf{Supplementary Figure }}
\caption{\textbf{Magnetization curves under different $\Gamma$ values.} 
The systems with the three $\Gamma/|K|$ ratio all 
exhibit two distinct field-induced phase transitions.
The low transition field $B_{\rm c1}$, marked by thick yellow bars, 
remains nearly constant around 35~T with varying $\Gamma/|K|$ values, 
indicating small $\Gamma$ dependence. 
In contrast, the high transition field $B_{\rm c2}$ (denoted by color-matched arrows)
decreases significantly with increasing $|\Gamma|/|K|$,
corresponding to the reduction in the range of the intermediate-field QSL regime. 
All curves are calculated using a unified energy scale of $|K| = 19$~meV. 
The lower three curves correspond to the magnetization results (left vertical axis), 
while the upper ones show their field derivatives (right vertical axis).
}
\label{FigS:Para}
\end{figure*}

As there are remarkable similarities in the finite-temperature 
and field-induced properties between $\alpha$-RuCl$_3$ and NCTO, 
it suggests that it might be a good starting point to adopt 
the model parameters of $\alpha$-RuCl$_3$. 
While the $\alpha$-RuCl$_3$ parameter is constructed based on 
$K$-$J_{1,2,3}$-$\Gamma$-$\Gamma'$ model
with ferromagnetic (FM) Kitaev interactions~\cite{Han2021}, 
we first perform a self-dual transformation~\cite{Chaloupka2015} on the $\alpha$-RuCl$_3$ model parameters~\cite{Han2021} 
to obtain an effective description with AFM Kitaev interactions, 
yielding a parameter set with $\Gamma/|K| = -0.3772$.
Subsequently, we find that
increasing the value of $|\Gamma|/|K|$ 
leads to a decrease in the high transition field strength $B_{\rm c2}$ and thus a
narrowing of the intermediate-field QSL regime, 
as shown in {Supplementary Fig.~\ref{FigS:Para}},
in better agreement with recent high-field magnetization measurements~\cite{Zhou2024arXiv} 
of NCTO along the out-of-plane direction.
Through tuning the $\Gamma/|K|$ ratio to $-0.65$, 
we could simultaneously reproduce both the specific heat (see the main text Fig.~\B{1}\textbf{a})
and field-induced properties of NCTO. 
Furthermore, we have verified that this parameter set 
maintains the zigzag AFM order as the ground state
with the corresponding spin correlations and magnetic moments data
as presented in Supplementary Fig.~\ref{FigS1:SqMz}\textbf{i} and \textbf{i'}.
Such parameter set has also been included in Supplementary Table.~\ref{Tab:Parameters1}.

\begin{figure*}[h]
\includegraphics[angle=0,width=0.99\linewidth]{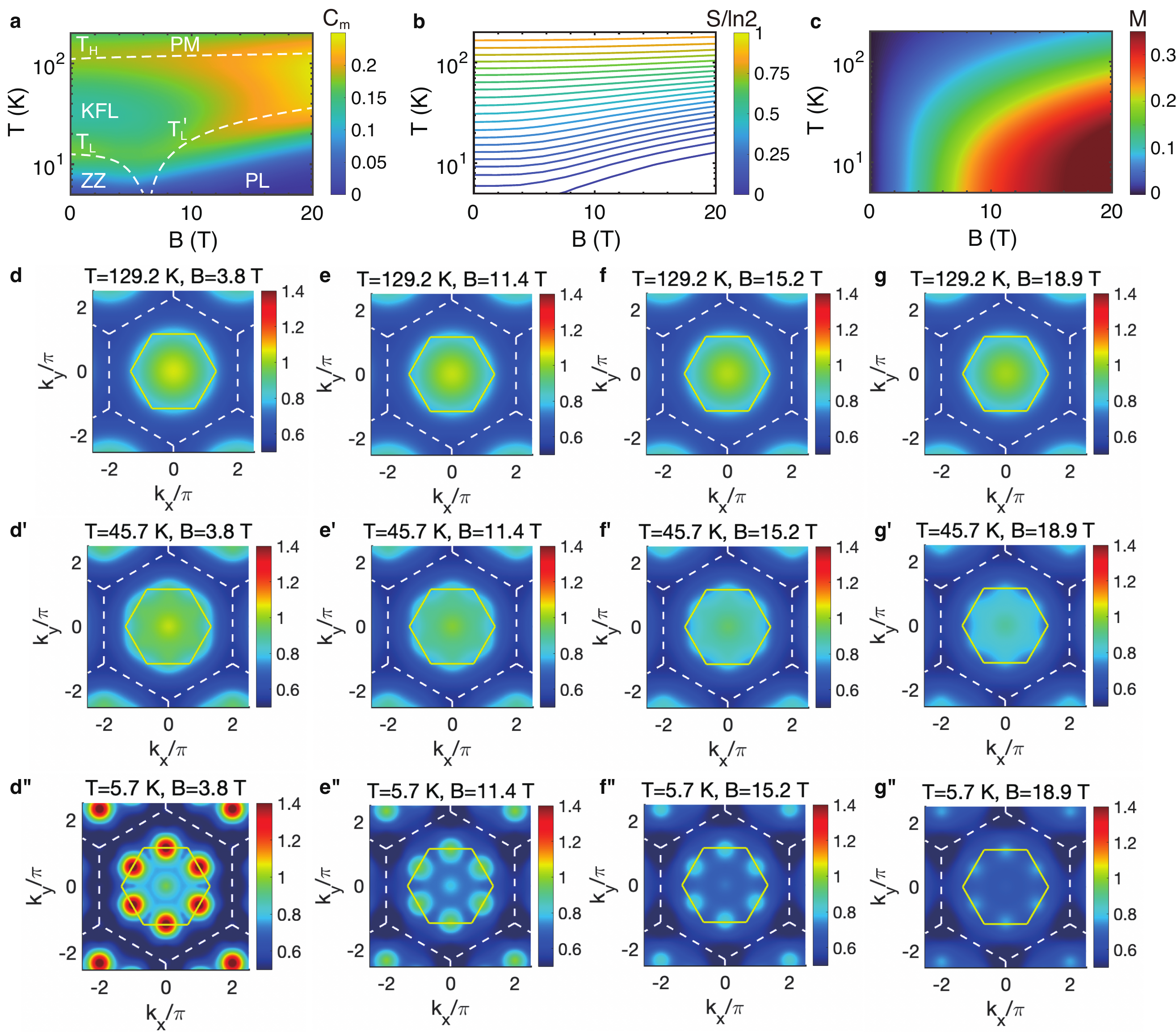}
\renewcommand{\figurename}{\textbf{Supplementary Figure }}
\caption{\textbf{Thermodynamic properties of NCTO model under in-plane magnetic fields.} 
\textbf{a-c} The landscapes of \textbf{a} specific heat $C_{\rm m}$, 
\textbf{b} thermal entropy $S/\rm{ln}2$, and \textbf{c} magnetization $M$ results 
with fields along $a^*$ axis ($B_{[l\ m\ n]} = B_{[-1\ 0\ 1]}$).  
The white dashed lines in panel \textbf{a} indicate the temperature scales 
$T_{\rm H}$, $T_{\rm L}$ and $T_{\rm L}'$, which also serves as schematic phase boundaries
separating paramagnetic (PM), zigzag (ZZ), Kitaev fractional liquid (KFL) and polarized (PL) phases.
\textbf{d-g} Calculated spin structure factors ${\tilde S}(\textbf{q})$ (see the main text for definition) 
under various magnetic fields at $T=129.2$~K, 
a temperature above the high-temperature scale $T_{\rm H}$. 
\textbf{d'-g'} Structure factors at an intermediate temperature $T=45.7$~K.
The ``M-star'' features in panels (\textbf{d'-g'})
signify the persistence of the KFL regime over a finite field range. 
\textbf{d''-g''} Structure factors at $T=5.7$~K.
The intensity at the M-point in the Brillouin zone is suppressed 
after the phase transition occurs with increasing in-plane fields.
}
\label{FigS:InPlane}
\end{figure*}

\subsection{Finite-temperature properties of NCTO model under in-plane fields}
\label{SecSM:InPlane}

\textbf{In-plane field-induced phase transitions.}
The magnetization process of NCTO under in-plane $a^*$ direction magnetic fields 
has been a focus in recent experimental studies~\cite{Yao2020, Lin2021NC, LinG2024, Zhang2023PRB, Xiao2022, Hong2024}. 
Those works have proposed several scenarios, 
including: (a) a first-order transition from the (canted) zigzag-ordered AFM phase 
to a coexisting zigzag and spin-disordered phase at about 6~T, 
followed by subsequent transitions (around 8~T) to a spin-disordered phase (possibly a QSL)
and finally to a spin-polarized phase (around 10~T)~\cite{LinG2024, Lin2021NC}; 
(b) a first-order phase transition at about 6~T and then continuously to the polarized phase~\cite{Takeda2022PRR}; 
or (c) a first-order transition from the triple-Q phase at about 8~T, 
and a quantum critical point (around 10~T)~\cite{Hong2024}.

In this work, employing the XTRG and DMRG methods 
with our proposed model parameters as listed in Supplementary Table.~\ref{Tab:Parameters1}, 
we have calculated the specific heat and thermodynamic entropy 
under in-plane magnetic fields up to 20~T. 
Our two-dimensional numerical simulations reveal 
only a single phase transition (no intermediate QSL phase) 
from the ordered phase to a (partially) polarized phase. 
Notably, the aforementioned first-order transition from 
``the (canted) zigzag or triple-Q ordered phase'' 
to a ``canted AFM'' or a ``coexisting zigzag and spin-disordered phase''
--- which possibly originates from three-dimensional material properties
--- remains undetectable in our two-dimensional finite-size study. 

\textbf{The landscapes of specific heat, thermal entropy, and magnetizations.}
Supplementary Fig.~\ref{FigS:InPlane}\textbf{a} presents a landscape of specific heat $C_{\rm m}$.
The temperature scale $T_{\rm H}$ associated with 
the high-temperature specific heat peak 
remains nearly unchanged across varying magnetic fields, 
while the low-temperature peak exhibits a non-monotonic response. 
Specifically, the low-temperature characteristic scale 
(marked as $T_{\rm L}$ in Supplementary Fig.~\ref{FigS:InPlane}\textbf{a})
initially shifts toward lower temperatures with weakening intensity as the field increases, 
but undergoes a notable reversal 
(marked as $T_{\rm L}'$ in Supplementary Fig.~\ref{FigS:InPlane}\textbf{a}) near 6 T
back to higher temperatures. 
This behavior suggests the emergence of a field-induced phase transition 
in the low-temperature regime.
The white dashed curves serving as guides to the eye, which indicates the phase boundaries
separating distinct phases in the phase diagram.

The thermodynamic entropy results also indicates the existence of a phase transition below 10~T, 
as evidenced in Supplementary Fig.~\ref{FigS:InPlane}\textbf{b}. At low magnetic fields ($B < 6$~T), the isentropic curves are densely distributed near $T_{\rm H}$ and $T_{\rm L}$, signify large entropy release near the two characteristic temperature scales, while sparsely distributed in the Kitaev fractional liquid (KFL) regime between these scales, and in the low-temperature zigzag ordered phase. Above 6~T, a marked upward curvature emerges in the isotropic curves,
demonstrating systematic entropy reduction in the low-temperature spin-polarized phase. 

We have also mapped the landscape of magnetization results $M = (-1 \langle S^x \rangle + 0 \langle S^y \rangle + 1 \langle S^z \rangle)/{\sqrt{2}}$ as shown in Supplementary Fig.~\ref{FigS:InPlane}\textbf{c}. The lower-right-corner regime indicates large magnetic moment,
which indicates that the system undergoes a slow convergence to the fully polarized phase.

\textbf{Spin correlations.}
In the lower part of Supplementary Fig.~\ref{FigS:InPlane}, we present the results of the spin-structure factors ${\tilde S}(\textbf{q})$ at different in-plane magnetic fields for three characteristic temperatures. It can be seen that when the temperature is relatively high ($T>T_{\rm H}$), as shown in Supplementary Fig.~\ref{FigS:InPlane}\textbf{d-g}, the structure factors change little with the magnetic field. At intermediate temperatures ($T_{\rm L}< T < T_{\rm H}$) and a lower field (see Supplementary Fig.~\ref{FigS:InPlane}\textbf{d'-g'}), we observe that the structure factor exhibits an M-star feature, which is consistent with the characteristics of the KFL phase~\cite{Li2020b, Han2021}. When the temperature is low ($T<T_{\rm L}$), and the magnetic field is small ($B <6$~T), i.e., Supplementary Fig.~\ref{FigS:InPlane}\textbf{d"}, there is a relatively large intensity at the M-point in the Brillouin zone, indicating that the system is in the zigzag-ordered phase; when the magnetic field is large ($B>10$~T, see Supplementary Fig.~\ref{FigS:InPlane}\textbf{e"-g"}), the intensity at the M-point is suppressed, indicating that the system enters the partially polarized phase. These characteristics of spin correlations, together with the specific heat, entropy, and magnetization mentioned above, enable us to obtain the finite-temperature phase diagram under in-plane magnetic fields as summarized in Supplementary Fig.~\ref{FigS:InPlane}\textbf{a}. 

\begin{figure*}[h!]
\includegraphics[angle=0,width=0.45\linewidth]{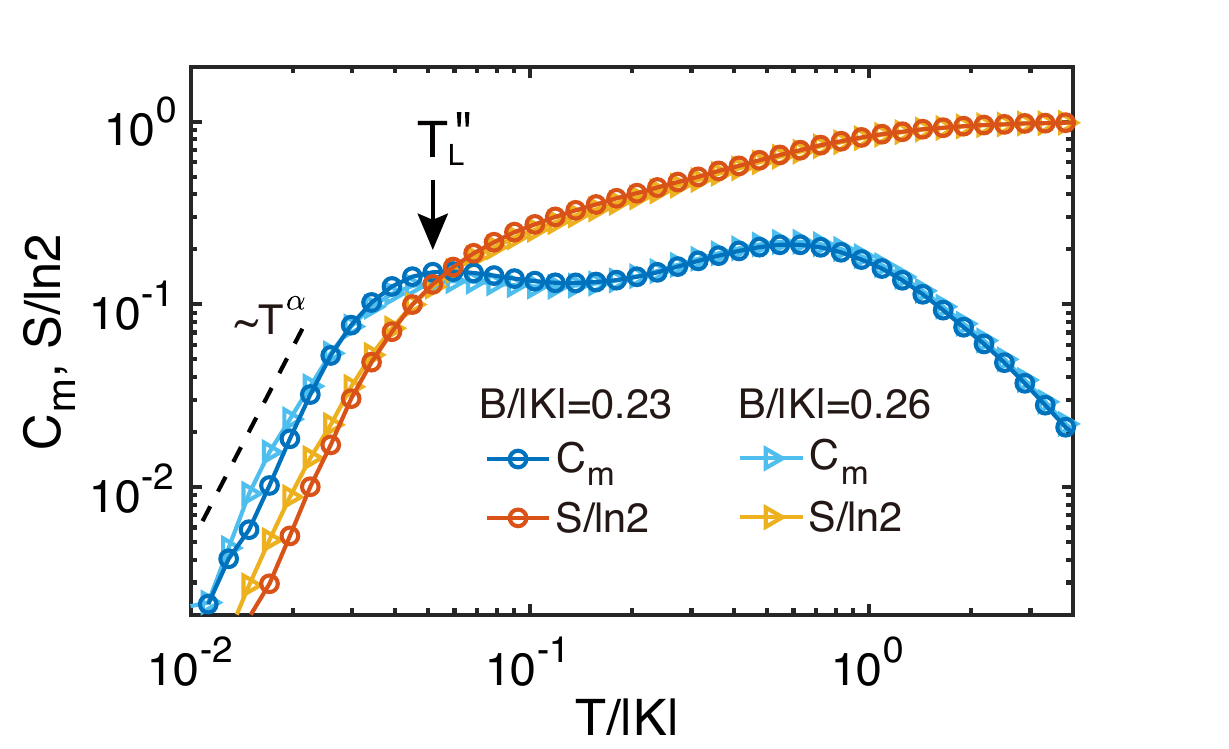}
\renewcommand{\figurename}{\textbf{Supplementary Figure }}
\caption{\textbf{Thermodynamic properties of NCTO model at fields $B/|K| \simeq 0.23$ and $0.26$.} 
The specific heat $C_{\rm m}$ and thermal entropy $S/{\rm ln}2$ curves 
with these two fields are plotted in logarithmic coordinates, 
which clearly demonstrate the power-law behavior $T^{\alpha}$
under the low-temperature scale $T_{\rm L}''$ (see the black dashed line).
}
\label{FigS:Intermediate}
\end{figure*}

\subsection{Additional results of field-induced gapless quantum spin liquid}
\label{SecSM:IGP}

\textbf{Low-temperature thermodynamic scaling in the intermediate-field gapless phase (IGP).} 
In addition to the low-temperature thermodynamic results at $B/|K|\simeq0.23$ presented in Fig.~\B{3}{\textbf{b}} of the main text, we have also calculated the specific heat and thermal entropy curves at a different field $B/|K|\simeq0.26$, also within the IGP, on YC4$\times$10$\times$2 lattices with $D=400$. The comparison of the results with these two fields are shown in Supplementary Fig.~\ref{FigS:Intermediate}. It is observed that the specific heat and entropy curves exhibit small variation between $B/|K|\simeq0.23$ and $0.26$. We note that a field difference of $0.03|K|$ in natural units corresponding to approximately 7~T in experimental units.
Below the low temperature scale $T_{\rm L}''$, the curves with $B/|K|\simeq0.26$ demonstrate nearly identical behavior to those at $B/|K|\simeq0.23$, both following the power-law scaling. 
This consistency strongly supports the gapless nature of the low-temperature intermediate-field phase.

\begin{figure*}[h]
\includegraphics[angle=0,width=0.95\linewidth]{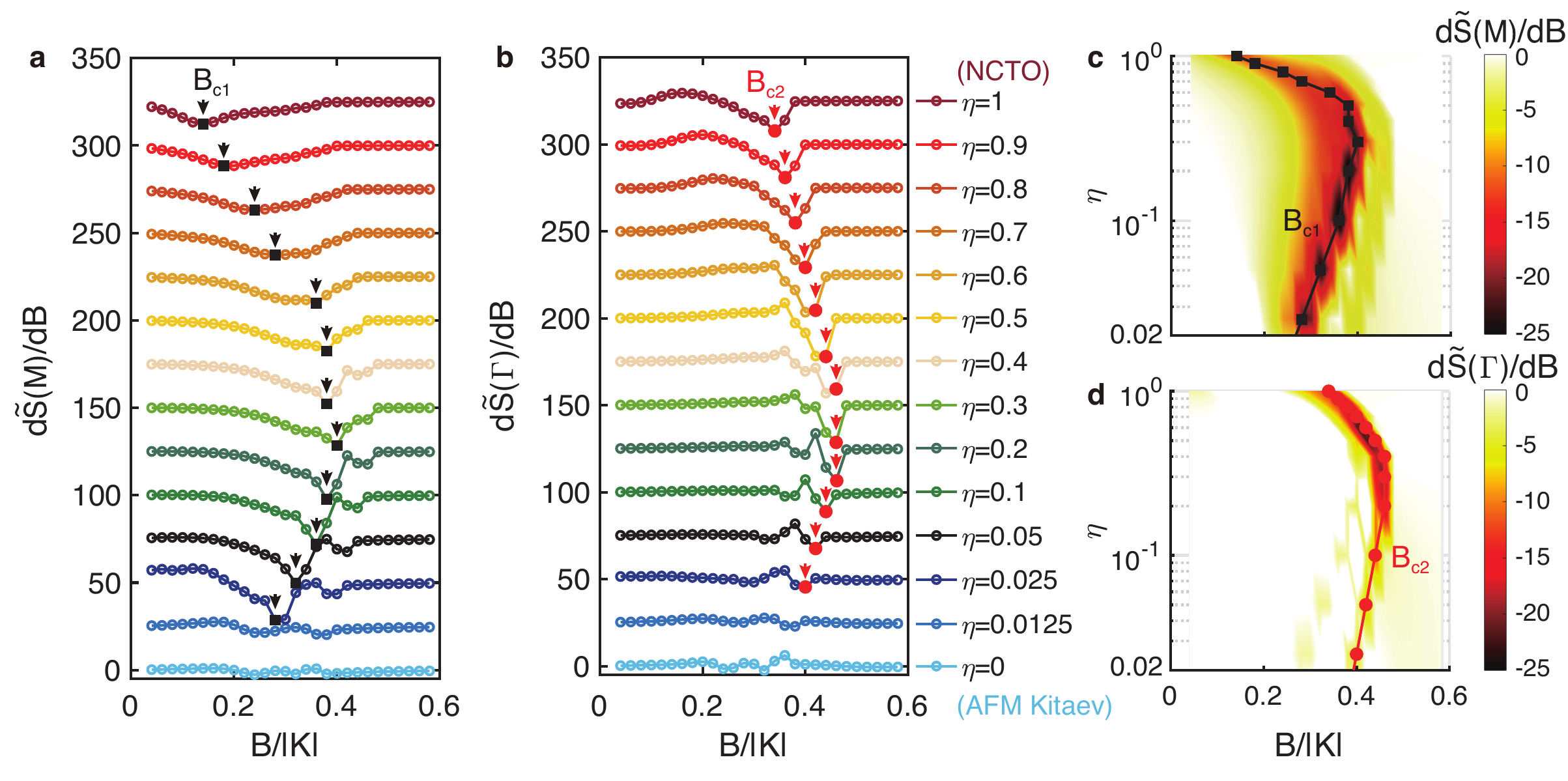}
\renewcommand{\figurename}{\textbf{Supplementary Figure }}
\caption{\textbf{Determination of the phase boundaries of IGP.} 
\textbf{a,b} The derivatives of the spin structure factors ${\rm d}\tilde{S}(M)/{\rm d}B$ and ${\rm d}\tilde{S}(\Gamma)/{\rm d}B$ versus out-of-plane magnetic field $B/|K|$. For $\eta \geq 0.025$, the dip minima in panels \textbf{a} and \textbf{b} (marked by black and red arrows, respectively) correspond to the transition fields $B_{\rm c1}$ and $B_{\rm c2}$. The curves in these plots have been vertically offset for clarity.
\textbf{c,d} The contour plots of ${\rm d}\tilde{S}(M)/{\rm d}B$ and ${\rm d}\tilde{S}(\Gamma)/{\rm d}B$ under different $B/|K|$ and $\eta$ values. The dark red regime corresponding to 
the dips of the curves in panels \textbf{a,b}, with the phase boundaries indicated by the black squares and red dots.
}
\label{FigS:dSqdB}
\end{figure*}

\textbf{Determination of the IGP phase boundaries.}
The boundaries of the IGP under an out-of-plane magnetic field for various $\eta$ values can be determined by analyzing the derivative of the spin structure factor. The system is located in the zigzag-ordered phase above certain $\eta$ values, characterized by the M-point structure factors $\tilde{S}(M)$. Therefore, the derivative ${\rm d}\tilde{S}(M)/{\rm d}B$, as shown in Supplementary Fig.~\ref{FigS:dSqdB}\textbf{a}, can be used to pin down the phase boundary. The positions of the dips in these curves with respect to the applied magnetic field correspond to the transition field $B_{\rm c1}$. These dip positions are indicated by black arrows and presented in Fig.~\B{3} and Fig.~\B{4} of the main text.

The $B_{\rm c2}$ for the phase transition from IGP to the polarized phase can be determined by the spin structure factor $\tilde{S}(\Gamma)$. In Supplementary Fig.~\ref{FigS:dSqdB}\textbf{b}, we present the ${\rm d}\tilde{S}(\Gamma)/{\rm d}B$ for various $\eta$ values, which reaches a minimum and subsequently converges rapidly to zero. The position of this dip (see the red arrows) determines the higher transition field $B_{\rm c2}$. 

In Supplementary Fig.~\ref{FigS:dSqdB}\textbf{c,d}, we show the contour plots of the ${\rm d}\tilde{S}(M)/{\rm d}B$ and ${\rm d}\tilde{S}(\Gamma)/{\rm d}B$ results. The red regimes in the plots indicate the dip positions and visualize the phase boundaries. These results indicate that the IGP in realistic NCTO model and pure AFM Kitaev model are adiabatically connected, i.e., they belong to the same phase. It thus provides strong evidence for the realization of Kitaev-derived spin liquid phase in the realistic Kitaev magnet NCTO.


\end{document}